\documentclass[letterpaper,10pt,xcolor=dvipsnames]{article}

\usepackage{style}
\usepackage[english]{babel} 
\usepackage[utf8]{inputenc}
\usepackage{mathrsfs}
\usepackage{amsmath,amssymb,amsfonts}
\usepackage{graphicx}
\usepackage{hyperref}
\usepackage{multirow}
\usepackage{multicol}
\usepackage{adjustbox}
\usepackage{cleveref}
\usepackage{dsfont}
\usepackage{xcolor}

\crefname{equation}{Eq.}{Eqs.}
\crefname{figure}{Fig.}{Figs.}
\crefname{section}{Sec.}{Secs.}
\crefname{table}{Tab.}{Tabs.}
\crefname{appendix}{Appx.}{Appx.}
\Crefname{equation}{Equation}{Equations}
\Crefname{figure}{Figure}{Figures}
\Crefname{section}{Section}{Sections}
\Crefname{table}{Table}{Tables}
\Crefname{appendix}{Appendix}{Appendices}

\numberwithin{equation}{section}
\DeclareMathAlphabet{\mathcal}{OMS}{cmsy}{m}{n}
\DeclareMathOperator{\atantwo}{atan2}

\begin{document}


\title{Diattenuation of Brain Tissue and its \\Impact on 3D Polarized Light Imaging}

\author{Miriam Menzel$^{1,}$*, Julia Reckfort$^{1,}$*, Daniel Weigand$^{2}$, Hasan Köse$^{1}$, Katrin Amunts$^{1,3}$, and Markus Axer$^{1}$}

\address{$^1$Institute of Neuroscience and Medicine (INM-1), Forschungszentrum Jülich, Jülich 52425, Germany\\
$^2$JARA-Institute for Quantum Information, RWTH Aachen University, Aachen 52056, Germany\\
$^3$Cécile and Oskar Vogt Institute of Brain Research, University of Düsseldorf, Düsseldorf 40204, Germany}

\email{* authors for correspondence (m.menzel@fz-juelich.de, j.reckfort@fz-juelich.de)}


\begin{abstract}
3D-Polarized Light Imaging (3D-PLI) reconstructs nerve fibers in histological brain sections by measuring their birefringence. This study investigates another effect caused by the optical anisotropy of brain tissue -- diattenuation. Based on numerical and experimental studies and a complete analytical description of the optical system, the diattenuation was determined to be below 4\,\% in rat brain tissue. It was demonstrated that the diattenuation effect has negligible impact on the fiber orientations derived by 3D-PLI. The diattenuation signal, however, was found to highlight different anatomical structures that cannot be distinguished with current imaging techniques, which makes Diattenuation Imaging a promising extension to 3D-PLI.

\vspace{0.5mm}

\center{\textbf{Keywords: polarized light imaging; diattenuation imaging; dichroism; polarization optics; transmittance; Müller-Stokes calculus; brain structure; white matter}}
\end{abstract}


\section{Introduction}
\label{sec:Introduction}

In order to understand the organization and function of the human brain, it is essential to study its fiber architecture, i.\,e. the spatial organization of the short- and long-range nerve fibers. Mapping this highly complex fiber architecture requires specific imaging techniques that resolve the  orientations of the fibers not only on a high spatial resolution but also on a large field of view of up to several centimeters.

The microscopy technique \textit{3D-Polarized Light Imaging (3D-PLI)} introduced by Axer et al.\ \cite{MAxer2011_1, MAxer2011_2} meets these specific requirements. It reveals the three-dimensional architecture of nerve fibers in sections of whole post-mortem brains with a resolution of a few micrometers. The orientations of the fibers are obtained by measuring the birefringence (axes of optical anisotropy) of unstained histological brain sections with a polarimeter. The measurement provides strong contrasts between different fiber structures and allows a label-free microscopy and reconstruction of densely packed myelinated fibers in human brains and those of other species. 

Birefringence of brain tissue is mainly caused by the regular arrangement of lipids and proteins in the myelin sheaths \cite{schmitt1939, martenson, koike-tani2013}. The optical anisotropy that causes birefringence (anisotropy of refraction) also leads to diattenuation (anisotropy of attenuation) \cite{chenault1993, mehta2013}. In diattenuating materials, the intensity of the transmitted light depends on the orientation of polarization of the incident light \cite{chipman1989, chenault1993, chipman}. 
If the diattenuation is solely caused by anisotropic absorption, it is typically called \textit{dichroism} \cite{ghosh2011, huang2006}. In the literature, diattenuation and dichroism are sometimes used as synonyms. Here, the term \textit{diattenuation} is used to describe the overall anisotropic attenuation of light that is caused not only by absorption but also by scattering.

As diattenuation leads to polarization-dependent attenuation of light, it might have an impact on the polarimetric measurement of 3D-PLI and consequentially affect the measured nerve fiber orientations. In this study, we investigated the diattenuation of brain tissue and its impact on the measured 3D-PLI signal for the first time.

Diattenuation as well as birefringence can be measured by conventional Müller-matrix polarimetry \cite{lu1996,chipman,ghosh2008,ghosh2010,ghosh2011} or by polarization-sensitive optical coherence tomography (PS-OCT) \cite{hee1992,pircher2011}. While PS-OCT uses the interference of the backscattered light to provide a depth profile of the sample, Müller polarimetry measures the intensity of the transmitted light under a certain angle. Often, incomplete Müller polarimeters are used that measure only the linear birefringence and diattenuation of a sample \cite{chenault1993, mehta2013}. In the present study, a combined measurement of (linear) birefringence and diattenuation was performed with an in-house developed polarimeter that analyzes the light transmitted through the sample \cite{MAxer2011_1,MAxer2011_2}.

Previous measurements that study the diattenuation of a sample were performed on non-biological phantoms (polarizing filters \cite{chenault1993,chen2009}, Siemens star \cite{mehta2013}) as well as on collagen \cite{swami2006}, tendon \cite{jiao2002,park2004,todorovic2004}, muscle \cite{park2004}, heart \cite{fan2013,todorovic2004}, skin \cite{jiao2003,makita2010,martin2013,smith2001,ghosh2011}, eye \cite{westphal2016,makita2010}, and biopsy tissue \cite{soni2013,wang2014} of animals or humans. Several studies investigated the diattenuation of the retinal nerve fiber layer (RNFL) \cite{huang2003_1,huang2003,naoun2005,huang2006} which only contains unmyelinated nerve fibers. To our knowledge, the diattenuation of myelinated nerve fibers and the diattenuation of brain tissue have not been addressed before and would need to be quantified.

The diattenuation of tissue reported in the above studies was much smaller than the birefringence of the investigated samples and mostly of secondary interest. As the diattenuation might influence the measured birefringence values, a couple of studies have been performed to estimate the error induced by diattenuation \cite{park2004,makita2010,mehta2013}. For the 3D-PLI measurement, the question arises in how far diattenuation influences the outcome of the measurement and  what are the consequences to the interpretation of the measured signal.  

In other studies, diattenuation has been used to quantify tissue properties (e.\,g.\ thickness \cite{naoun2005}, concentration of glucose \cite{westphal2016,ghosh2011}) and to distinguish between healthy and pathological tissue (cancerous tissue \cite{soni2013}, burned/injured tissue \cite{jiao2003,smith2001}, tissue from eye diseases \cite{naoun2005,huang2006}). Hence, diattenuation might also provide interesting structural information about the brain tissue and \textit{Diattenuation Imaging (DI)} could be a useful extension to 3D-PLI. 

The present study was therefore designed (a) to quantify the diattenuation of brain tissue, (b) to quantify the impact of diattenuation on the measured 3D-PLI signal, and (c) to investigate whether the diattenuation signal contains useful information about the brain tissue structure.  

\begin{figure}[h]
\centering
\fbox{\includegraphics[width=1 \textwidth]{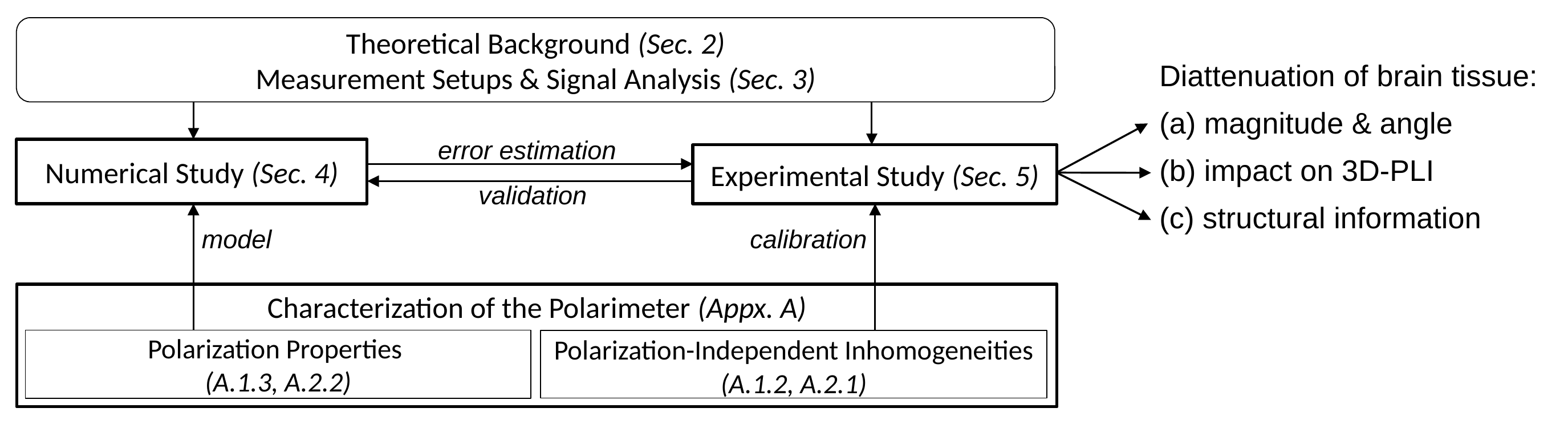}}
\caption{Design of the study.}
\label{fig:Outline}
\end{figure}

The study design is reflected in the structure of this paper (see \cref{fig:Outline}) which is composed of a numerical study (\cref{sec:Numerical_Study}) and an experimental study (\cref{sec:Experimental_Study_on_Brain_Tissue}). The numerical study was performed because the above literature suggests that the diattenuation signal is small and could also be caused by non-ideal optical components of the employed polarimeter. The numerical study estimates the impact of the non-ideal system parameters and the tissue diattenuation on the reconstructed fiber orientations and the measured diattenuation. In the experimental study, the determined error estimates were taken into account to quantify the diattenuation of brain tissue and its impact on 3D-PLI. The experimental study was performed exemplary on five sagittal rat brain sections. The numerical and experimental study are presented as separate studies, each divided in methods, results, and discussion. 

The analytical model used for the analysis of these studies is developed in \cref{sec:Measurement_Setups_Analysis}. The model considers not only the birefringence but also the diattenuation of brain tissue as well as non-ideal system components. The non-ideal polarization properties of the polarimeter used for the numerical study and the polarization-independent inhomogeneities used for calibrating the experimental measurements were characterized in a preliminary study presented in \cref{sec:Characterization_LAP}. 

In an overall discussion at the end of this paper (\cref{sec:Overall-Discussion}), the results of the experimental study are compared to the predictions of the numerical study to validate the developed model.
A list of all symbols and abbreviations used throughout this paper can be found in \cref{sec:Symbols}.

\section{Theoretical Background}
\label{sec:Theoretical_Background}

This section introduces the physical principles and the mathematical notation used in this study. 
Apart from birefringence and diattenuation\footnote{In this paper, the terms `birefringence' and `diattenuation' are used to describe linear birefringence and diattenuation. Their circular counterparts are neglected in this study because they are expected to contain not much information about the nerve fiber orientation and have no significant impact on the measured signals.}, the Müller-Stokes calculus \cite{mueller1943} is described which will be used in the following to derive analytical expressions for the measured light intensities.


\subsection{Birefringence}
\label{sec:Birefringence}

In optically anisotropic media, the refractive index depends on the direction of propagation and on the polarization state of the incident light. This anisotropic refraction, known as birefringence, can be caused by regular molecular structures, but also by orderly arranged units far larger than molecules \cite{born}.

Light that travels through a birefringent medium experiences a phase difference (\textit{retardance}) $\delta$ between two orthogonal polarization components (ordinary and extraordinary wave with refractive indices $n_o$ and $n_e$), which changes the state of polarization of the light.
For example, a quarter-wave retarder with $\delta=\pi/2$ transforms linearly polarized light into (right-/left-handed) circularly polarized light when its fast axis is oriented at an angle of (+/-) $45^{\circ}$ to the direction of polarization of the incident light \cite{born}. 

Previous studies have shown that myelinated nerve fiber bundles exhibit uniaxial negative birefringence ($n_e < n_o$) and that the optic axis (direction of optical anisotropy) is oriented in the direction of the fiber bundle \cite{goethlin1913,schmitt1939,vidal1980,menzel2015}. 
Like in all biological tissues, the birefringence of the nerve fibers is supposed to be small as compared to the refractive index of the fibers \cite{beuthan1996,ghosh2011}. In this case, the induced phase shift can be approximated as \cite{larsen2007,menzel2015}:
\begin{align}
\delta \approx \frac{2 \, \pi}{\lambda} \, d \, \Delta n \, \cos^2 \alpha, 
\label{eq:delta_approx}
\end{align}
where $\lambda$ is the wavelength of the light, $d$ the thickness of the medium, $\Delta n$ the birefringence, and $\alpha$ the out-of-plane angle of the optic axis (i.\,e. the inclination angle of the nerve fibers, cf.\ \cref{fig:Setups}h).


\subsection{Diattenuation}
\label{sec:Diattenuation}

Diattenuation refers to anisotropic attenuation of light which can be caused by absorption (dichroism) as well as by scattering \cite{born,ghosh2011,chipman1989}. In diattenuating materials, the transmitted light intensity depends on the polarization state of the incident light: The transmitted light intensity is maximal $(I_{\text{max}})$ for light polarized in a particular direction and minimal $(I_{\text{min}})$ for light polarized in the corresponding orthogonal direction. The diattenuation is defined as \cite{chipman, chenault1993}:
\begin{align}
D = \frac{I_{\text{max}} - I_{\text{min}}}{I_{\text{max}} + I_{\text{min}}}, \,\,\,\,\,\,\,\,\,\,\,\ 
0 \leq D \leq 1.
\label{eq:D}
\end{align}
The average transmittance, i.\,e.\ the fraction of unpolarized light that is transmitted through a sample, is given by \cite{chenault1993}:
\begin{align}
\tau = \frac{I_{\text{max}} + I_{\text{min}}}{2\,I_0}, \,\,\,\,\,\,\,\,\,\,\,\ 
0 \leq \tau \leq 1,
\label{eq:tau}
\end{align}
with $I_0$ being the intensity of the incident light. 
Optical elements with high diattenuation are used to create linearly polarized light. 
An ideal linear diattenuator (polarizer) fulfills $D=1, \tau = 1/2$, i.\,e.\ the intensity of unpolarized light is reduced by one half. 

As diattenuation and birefringence are usually caused by the same anisotropic structure, the principal axes of diattenuation are assumed to be coincident with the principal axes of birefringence \cite{chenault1993}. 
In this case, dichroism (anisotropic absorption) and birefringence (anisotropic refraction) can be described by the imaginary and real parts of a complex retardance \cite{born,fan2013}. Thus, diattenuation caused by dichroism (no scattering) is approximately proportional to $\delta$.


\subsection{Müller-Stokes Calculus}
\label{sec:Muller-Stokes_Calculus}

The Müller-Stokes calculus allows a complete mathematical description of polarized light. It is also suitable for partially polarized and incoherent light. The polarization state of light is described by a $4 \times 1$ Stokes vector and the optical elements of the polarimetric setup are described by $4 \times 4$ Müller matrices.

\paragraph{Stokes vectors:} 

The Stokes vector $\vec{S}$ is defined in spherical coordinates as \cite{collett}:
\begin{align}
\vec{S} = 
\begin{pmatrix} I \\
				I\,p\,\cos({2\psi})\,\cos({2\chi}) \\
				I\,p\,\sin({2\psi})\,\cos({2\chi}) \\
				I\,p\,\sin({2\chi})
\end{pmatrix}, \,\,\,\,\,\,\, 
p = \frac{\sqrt{S_1^2 + S_2^2 + S_3^2}}{S_0}, 
\label{eq:StokesVector2}
\end{align}
where $I$ is the total intensity of the light beam, $p \in [0,1]$ is the degree of polarization, and $\psi \in [0, \pi]$ and $\chi \in [-\pi/4, \pi/4]$ are the spherical angles which determine the orientation of the vector ($S_1$, $S_2$, $S_3$) in the Poincar\'{e} sphere, i.\,e. the (linear/circular) polarization of the light. 
For completely unpolarized light, the degree of polarization is zero ($p=0$) and the Stokes vector simplifies to: $\vec{S}_{\text{unpol}} = (I,0,0,0)^{\top}$.

\paragraph{Müller matrices:}
The optical elements used in this study can be represented by a wave retarder and/or diattenuator. The Müller matrix for a general wave retarder and diattenuator (with retardance $\delta$, diattenuation $D$, and average transmittance $\tau$) is given by \cite{chipman, chenault1993}:
\begin{align}
\mathcal{M}(\delta,\,D,\,\tau) 
= \tau
\begin{pmatrix} 1 		& D		& 0 									& 0 \\
				D 		& 1		& 0									& 0 \\
				0 		& 0 	 	& \sqrt{1 - D^2} \,\, \cos\delta 	& \sqrt{1 - D^2} \,\, \sin\delta \\
				0 		& 0		& -\sqrt{1 - D^2} \,\, \sin\delta	& \sqrt{1 - D^2} \,\, \cos\delta
\end{pmatrix}, 
\label{eq:M}
\end{align}
where the principal axes of birefringence and diattenuation are coincident and aligned with the x- and y-axes of the reference frame (the fast axis of the retarder and the axis of \textit{maximum} intensity transmittance are aligned with the x-axis). This definition will be used in all subsequent formulas and derivations. To describe materials in which the transmitted light intensity is \textit{minimal} for polarizations along the x-axis, the variable $D$ in the above matrix needs to be replaced by the variable $(-D)$.

A rotation in counter-clockwise direction by an angle $\xi$ is described by the rotation matrix:
\begin{align}
R(\xi) = 
\begin{pmatrix} 1 		& 0				& 0 				& 0 \\
				0 		& \cos(2\xi)	& -\sin(2\xi)	& 0 \\
				0 		& \sin(2\xi) 	& \cos(2\xi) 	& 0 \\
				0 		& 0				& 0				& 1
\end{pmatrix}. 
\label{eq:R}
\end{align}

The Müller matrix of a retarder and/or diattenuator rotated in counter-clockwise direction by an angle $\xi$ is given by:
\begin{align}
\mathcal{M}(\xi,\,\delta,\,D,\,\tau) = R(\xi) \cdot \mathcal{M}(\delta,\,D,\,\tau) \cdot R(-\xi). 
\label{eq:M_rot}
\end{align}

When an input beam described by a Stokes vector $\vec{S}$ passes an assembly of optical elements described by a Müller matrix $\mathcal{M}$, the resulting output beam is given by:
\begin{align}
\vec{S}' = \mathcal{M} \cdot \vec{S},
\end{align}
where the first entry of the resulting Stokes vector ($S'_0$) describes the intensity of the output beam.


\section{Measurement Setups and Signal Analysis}
\label{sec:Measurement_Setups_Analysis}

\enlargethispage{0.5cm}

\begin{figure}[htbp]
\centering
\fbox{\includegraphics[width=1 \textwidth]{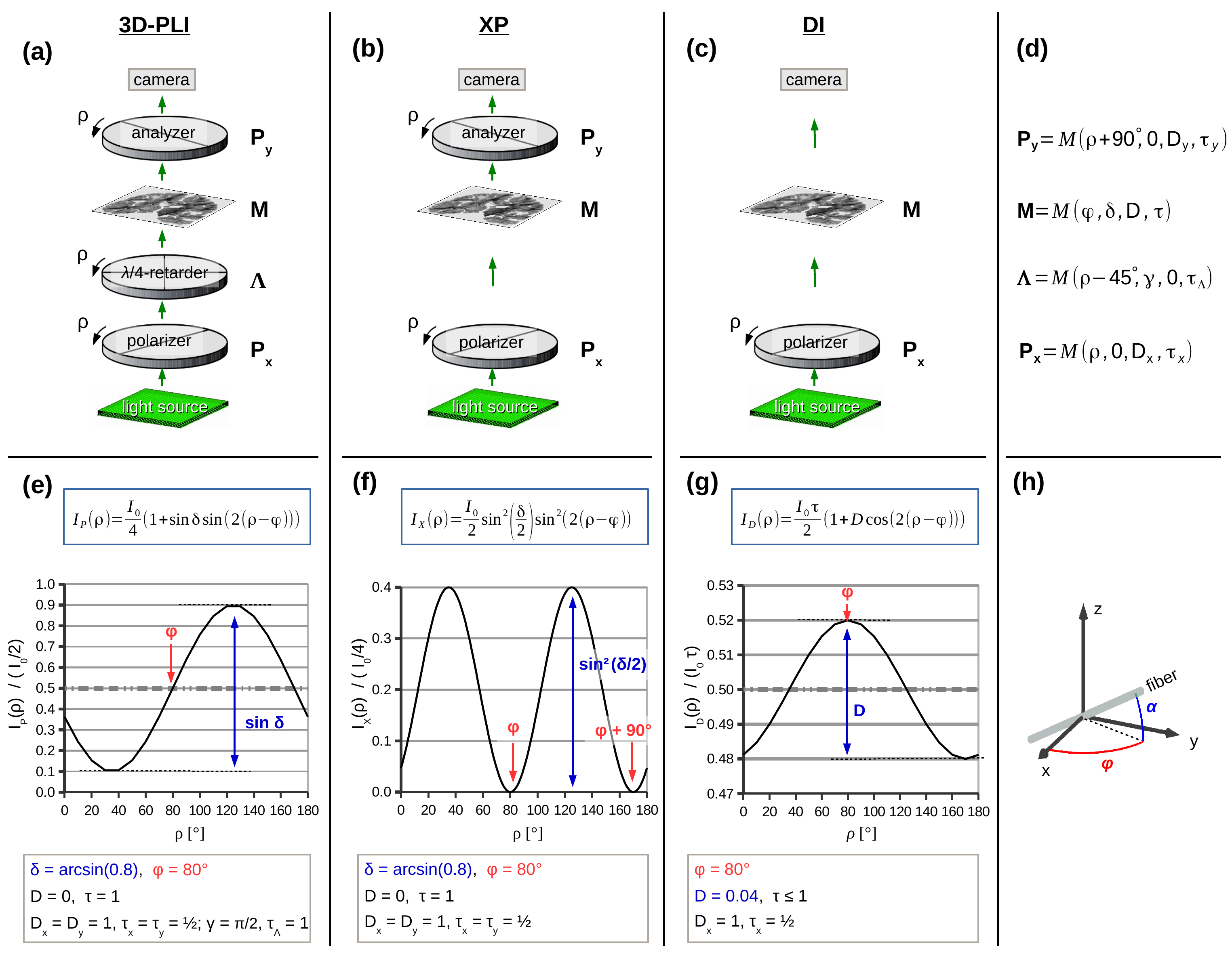}}
\caption{\textbf{(a-c)} Schematic of the setups for the 3D-PLI, XP, and DI measurements: For the 3D-PLI measurement (a), the brain section is placed between a pair of crossed linear polarizers (polarizer/analyzer) and a quarter-wave retarder. For the XP measurement (b), only the crossed linear polarizers are used while for the DI measurement (c) only the polarizer is used. For all measurement setups, the employed filters are rotated simultaneously by discrete rotation angles $\rho$ around the stationary specimen. \textbf{(d)} The transmitted light intensity is calculated using the Müller-Stokes calculus, in which each optical element is represented by a Müller matrix $\mathcal{M}(\xi, \delta, D, \tau)$ as defined in \cref{eq:M,eq:M_rot,eq:R}. \textbf{(e-g)} Analytically computed normalized light intensity profiles for the different measurement setups, assuming a retardance of $\delta = \arcsin(0.8)$, a fiber direction of $\varphi = 80^{\circ}$, and ideal filter properties ($D_x = D_y = 1$, $\tau_x = \tau_y = 1/2$; $\gamma = \pi/2$, $\tau_{\Lambda} = 1$): For the 3D-PLI and XP measurements, the diattenuation and absorption of the brain tissue were neglected ($D=0$, $\tau=1$). For the DI measurement, the tissue diattenuation was assumed to be $D=4\,\%$. The phase $\varphi$ of the intensity profiles (in red) is a measure for the in-plane fiber direction, while the amplitude (in blue) is correlated to the out-of-plane fiber inclination. \textbf{(h)} The three-dimensional fiber orientation is defined in spherical coordinates by the in-plane direction angle $\varphi$ and the out-of-plane inclination angle $\alpha$.}
\label{fig:Setups}
\end{figure}

To study the diattenuation of brain tissue and its impact on the measured 3D-PLI signal, unstained histological brain sections were measured with different polarimetric setups: 3D-Polarized Light Imaging (3D-PLI), crossed polars (XP) measurement, and Diattenuation Imaging (DI) (see \cref{fig:Setups}). The XP measurement was used as a reference for the 3D-PLI and DI measurements to estimate the impact of diattenuation on the measured birefringence signal.

The polarimeter used for the 3D-PLI, XP, and DI measurements\footnote{In previous publications, the polarimeter was referred to as \textit{Large-Area Polarimeter (LAP)} \cite{MAxer2011_1,MAxer2011_2,menzel2015,reckfort2015}.} consists of an LED light source, a linear polarizer (called \textit{polarizer}), a quarter-wave retarder, a specimen stage, a second linear polarizer (called \textit{analyzer}), and a CCD camera (see \cref{fig:Setups}a). The principal axis of the analyzer is oriented at $90^{\circ}$ with respect to the principal axis of the polarizer and the fast axis of the quarter-wave retarder is oriented at $-45^{\circ}$. The optical components of the polarimeter are described in \cref{sec:Components_LAP} in more detail. 

During a measurement, the filters (polarizer/retarder/analyzer) are rotated simultaneously in counter-clockwise direction by discrete rotation angles $\rho$. 
To realize the different measurement setups (3D-PLI, XP, DI), one or more filters are removed from the light path (cf.\ \cref{fig:Setups}a-c). 

For each measurement setup, an analytical expression of the transmitted light intensity was computed using the Müller-Stokes calculus. 
For this purpose, the optical components of the polarimeter were described by the Müller matrix $\mathcal{M}\,(\xi, \delta, D, \tau)$ of a rotating wave retarder and/or diattenuator as defined in \cref{eq:M,eq:M_rot,eq:R}. 
To account for non-ideal optical properties of the filters (see \cref{sec:Characterization_LAP}), the polarizer and the analyzer were considered to be general diattenuators with rotation angles $\rho$ and $(\rho+90^{\circ})$, diattenuations $D_x$ and $D_y$, and average transmittances $\tau_x$ and $\tau_y$, respectively. The quarter-wave retarder was considered as general retarder with rotation angle $(\rho-45^{\circ})$, retardance $\gamma$, and average transmittance $\tau_{\Lambda}$. 
The retardance of the linear polarizers and the diattenuation of the retarder were not included in this model because the filter measurements in \cref{sec:Characterization_Pol-Prop,sec:Polarization_Properties} have shown that they are negligible.

To account for birefringence as well as for a possible diattenuation of brain tissue, the brain section was described by the Müller matrix of a general wave retarder and diattenuator with retardance $\delta$, diattenuation $D$, and average transmittance $\tau$. The fast axis of the retarder and the axis of maximum intensity transmittance were both assumed to be oriented along the fiber axis.\footnote{The assumption that the axis of maximum intensity transmittance is oriented along the fiber axis is arbitrary. If the axis of \textit{minimum} intensity transmittance is oriented along the fiber axis, the following considerations are still valid when replacing the variable $D$ by the variable ($-D$).} The three-dimensional fiber orientation is defined in spherical coordinates by the in-plane direction angle $\varphi$ and the out-of-plane inclination angle $\alpha$ (cf.\ \cref{fig:Setups}h). 
Note that the Müller matrix describes only the net effect of the brain tissue and that the parameters ($\varphi$, $\delta$, $D$, $\tau$) do not necessarily correspond to the local tissue properties.

With the above definitions, the Müller matrices $\mathcal{M}\,(\xi, \delta, D, \tau)$ for the optical components read:
\begin{align}
&\text{Polarizer:}\quad\quad\,\, \,P_x(\rho,\,D_x,\,\tau_x) \,\,\,\,\,\,\,\equiv \,\,\mathcal{M}\,(\rho, 0, D_x, \tau_x), 
\label{eq:Polarizer_non-ideal} \\
&\text{Retarder:}\,\,\,\,\quad\quad \Lambda\,(\rho,\gamma,\tau_{\Lambda}) \,\,\,\,\,\,\, \,\,\,\,\,\,\,\equiv\,\, \mathcal{M}\,(\rho - 45^{\circ}, \gamma, 0, \tau_{\Lambda}), 
\label{eq:Retarder_non-ideal}\\
&\text{Brain Tissue:} \,\,\,\,\,\, M\,(\varphi,\,\delta,\,D,\,\tau) \,\,\,\,\,\, \equiv \,\,\mathcal{M}\,(\varphi, \delta, D, \tau), 
\label{eq:BrainTissue_non-ideal}\\ 
&\text{Analyzer:} \,\,\,\,\quad\,\,\,\,\, P_y\,(\rho,\,D_y,\,\tau_y) \,\,\,\,\,\,\equiv \,\,\mathcal{M}\,(\rho + 90^{\circ}, 0, D_y, \tau_y).
\label{eq:Analyzer_non-ideal}
\end{align}

The transmitted light intensity was computed by multiplying the above matrices and evaluating the first entry of the resulting Stokes vector.
To make the analytical expressions computable, the light emitted by the LED was assumed to be completely unpolarized ($\vec{S}_L = \vec{S}_{\text{unpol}}$) and the camera to be polarization-insensitive ($\vec{S}_c = \vec{S}_{\text{unpol}}$). 

The following sections describe the setups and signal analysis for the 3D-PLI, XP, and DI measurements.
All parameters derived from the 3D-PLI measurement were denoted by an index $P$, all parameters derived from the XP measurement by an index $X$, and all parameters derived from the DI measurement by an index $D$.


\subsection{Three-Dimensional Polarized Light Imaging (3D-PLI)}
\label{sec:3D-PLI}

For the 3D-PLI measurement, all filters of the polarimeter are used (see \cref{fig:Setups}a). This setup allows to measure the retardance of the birefringent brain sections and to derive the spatial orientation of the nerve fibers \cite{MAxer2011_1,MAxer2011_2}. The polarizer and the quarter-wave retarder transform the unpolarized light emitted by the light source into circularly polarized light which is then transformed into elliptically polarized light by the birefringent brain tissue. The amount of light that is transmitted through the analyzer depends on the orientation of the analyzer axis with respect to the optic axis of the brain tissue, indicating the predominant fiber orientation \cite{menzel2015}. 

The fiber orientation ($\varphi$,$\alpha$) is computed for each image pixel from the phase and amplitude of the corresponding sinusoidal light intensity profile that is obtained when plotting the transmitted light intensity $I(\rho)$ against the rotation angle $\rho$ of the filters (see \cref{fig:Setups}e). An analytical description of the transmitted light intensity can be derived by multiplying the matrices defined in \cref{eq:Analyzer_non-ideal,eq:Polarizer_non-ideal,eq:Retarder_non-ideal,eq:BrainTissue_non-ideal} and evaluating the first entry of the resulting Stokes vector ($S'_{0P}$):
\begin{align}
\vec{S}'_P(\rho) &= P_y\,(\rho,\,D_y,\,\tau_y) \cdot M\,(\varphi,\,\delta,\,D,\,\tau) \cdot  \Lambda\,(\rho, \gamma, \tau_{\Lambda}) \cdot P_x\,(\rho,\,D_x,\,\tau_x) \cdot \vec{S}_{\text{unpol}} \\
\Rightarrow \,\,\, I_P(\rho) &= \tau \, \tau_x \, \tau_y \,  \tau_{\Lambda} \, I_0  \, 
\bigg[ 1 
+ D_x\,D_y\, \sin\gamma \, \sin\delta\, \sqrt{1-D^2}\,\sin\big(2(\rho-\varphi)\big) \notag \\
& \quad\quad - D_x\,D_y\,\cos\gamma \, \left(\cos^2\big(2(\rho-\varphi)\big) + \sqrt{1-D^2}\,\cos\delta\,\sin^2\big(2(\rho-\varphi)\big) \right) \notag \\
& \quad\quad + D \, \big(D_x\,\cos\gamma - D_y\big) \cos\big(2(\rho-\varphi)\big)
\bigg],
\label{eq:IntensitySignal}
\end{align}
where $I_0$ denotes the intensity of the light source. Performing a discrete harmonic Fourier analysis on the acquired intensity signal yields:
\begin{align}
I_P(\rho) &= a_{0P} + a_{2P} \, \cos(2\rho) + b_{2P} \, \sin(2\rho) + a_{4P} \, \cos(4\rho) + b_{4P} \, \sin(4\rho), \\
a_{0P} &= \tau \, \tau_x \, \tau_y \, \tau_{\Lambda} \, I_0  \, \left( 1 - \frac{1}{2} D_x \, D_y \, \cos\gamma \, \left(1 + \sqrt{1-D^2} \, \cos\delta \right) \right), 
\label{eq:a0}\\
a_{2P} &= \tau \, \tau_x \, \tau_y \, \tau_{\Lambda} \, I_0 \, \left(D\,\big(D_x \, \cos\gamma - D_y\big)\,\cos(2\varphi) - \sqrt{1-D^2} \,\, D_x \, D_y \, \sin\gamma \,\, \sin\delta \, \sin(2\varphi) \right), 
\label{eq:a2}\\
b_{2P} &= \tau \, \tau_x \, \tau_y \, \tau_{\Lambda} \, I_0 \, \left(D\,\big(D_x \, \cos\gamma - D_y\big)\,\sin(2\varphi) + \sqrt{1-D^2} \,\, D_x \, D_y \, \sin\gamma \,\, \sin\delta \, \cos(2\varphi) \right), 
\label{eq:b2}\\
a_{4P} &= - \frac{1}{2} \, \tau \, \tau_x \, \tau_y \, \tau_{\Lambda} \, I_0 \, D_x \, D_y \, \cos\gamma \, \left( 1 - \sqrt{1-D^2}\,\,\cos\delta \right) \cos(4\varphi), 
\label{eq:a4}\\
b_{4P} &= - \frac{1}{2} \, \tau \, \tau_x \, \tau_y \, \tau_{\Lambda} \, I_0 \, D_x \, D_y \, \cos\gamma \, \left( 1 - \sqrt{1-D^2}\,\,\cos\delta \right) \sin(4\varphi). 
\label{eq:b4}
\end{align}

In the standard 3D-PLI analysis \cite{MAxer2011_1, MAxer2011_2, menzel2015}, the spatial fiber orientations ($\varphi$, $\alpha$) are derived from the measured Fourier coefficients assuming ideal linear polarizers ($D_x = D_y = 1$, $\tau_x = \tau_y = 1/2$), an ideal quarter-wave retarder ($\gamma = \pi/2, \tau_{\Lambda} = 1$), and no diattenuation of the brain tissue ($D = 0$). 
In this ideal case, the Fourier coefficients of fourth order vanish and \cref{eq:IntensitySignal} simplifies to: $I_P(\rho) = I_0 \, \big(1 + \sin\delta \, \sin(2(\rho-\varphi))\big) / 4$.

The Fourier coefficients of zeroth and second order are used to compute different parameter maps: the \textit{transmittance} ($I_{T,P}$), the \textit{direction} ($\varphi_P$), and the \textit{retardation} ($r_P$). 
The transmittance map represents the average transmitted light intensity over all rotation angles and is computed from the Fourier coefficient of zeroth order:
\begin{align}
I_{T,P} = 2 \, a_{0P}.
\label{eq:transmittance}
\end{align}
The direction describes the in-plane orientation angle of the fibers and is computed from the phase of the intensity profile (\cref{fig:Setups}e, in red):
\begin{align}
\varphi_P = \frac{\atantwo(-a_{2P}, b_{2P})}{2} \,\,\,\, \in \,\, [0, \pi).
\label{eq:direction}
\end{align}
The retardation is computed from the peak-to-peak amplitude of the intensity profile normalized by the transmittance $I_P(\rho)/I_{T,P}$ (\cref{fig:Setups}e, in blue):
\begin{align}
r_P \equiv \vert \sin \delta_P \vert = \frac{\sqrt{a_{2P}^2 + b_{2P}^2}}{a_{0P}},
\label{eq:retardation}
\end{align}
and is used to derive the inclination angle $\alpha_P$ of the fibers by applying \cref{eq:delta_approx} ($\delta_P \propto \cos^2\alpha_P$).

According to \cref{eq:delta_approx}, the phase shift $\delta$ scales with the material thickness, the birefringence, and the illumination wavelength. As these parameters cannot be determined separately, the measured retardation $r_P$ is normalized by the maximum measurable retardation $r_{\text{max}}$, which is assumed to correspond to a region that is completely filled with horizontal (birefringent) fibers \cite{MAxer2011_1}. From the normalized retardation, a modified inclination angle is computed:
\begin{align}
\tilde{\alpha}_P = \arccos \left( \sqrt{\frac{\arcsin(r_P)}{\arcsin(r_{\text{max}})}} \right).
\label{eq:alpha_corr}
\end{align}

Note that the above derivation of the fiber inclination angle assumes that the investigated brain region contains parallel fibers with similar tissue composition. This assumption does not affect the validity of the analytical model and the predictions of the numerical study in \cref{sec:Numerical_Study}. The experimental study in \cref{sec:Experimental_Study_on_Brain_Tissue} focuses only on the fiber direction and retardation. Extracting the fiber inclination in inhomogeneous brain regions is subject of current research and previous publications \cite{reckfort,wiese2014,wiese,dohmen2015}.

The intensity signal $I_P(\rho)$ is a measure of the fiber orientation as defined in \cref{fig:Setups}h. 
However, if the polarization properties of the filters are non-ideal or if the diattenuation of the brain tissue is non-zero, the measured fiber orientations ($\varphi_P$, $\alpha_P$) will not exactly correspond to the actual fiber orientations ($\varphi$, $\alpha$) because the measured Fourier coefficients ($a_{0P}$, $a_{2P}$, $b_{2P}$) depend on $D_x$, $D_y$, $\gamma$, and $D$.\footnote{In principle, the actual fiber direction angle $\varphi$ could be derived from ($b_{4P}/a_{4P}$). However, this is not feasible because $a_{4P}$ and $b_{4P}$ are much smaller than $a_{2P}$ and $b_{2P}$ ($\cos\gamma \ll 1$, for a quarter-wave retarder with $\gamma \approx \pi/2$), resulting in a much lower signal-to-noise ratio.} In the numerical study (\cref{sec:3DPLI_Numerical}), this dependency will be investigated in more detail. If the values for $D_x$, $D_y$, $\gamma$, and $D$ were known pixel-wise, the actual fiber orientation could exactly be computed from the above Fourier coefficients (see \cref{sec:Corr_Fiber_Orientation}). 


\subsection{Crossed Polars (XP) Measurement}
\label{sec:Crossed_Polars_Measurement}

The crossed polars (XP) measurement \cite{HAxer2011} allows the determination of the direction angle independently from $D_x$, $D_y$, $\gamma$, or $D$ and can therefore be used as a reference for the actual fiber direction $\varphi$. 
However, the XP measurement cannot replace 3D-PLI because the direction angle can only be determined in a value range of $[0^{\circ},90^{\circ})$ and the measurement gives no information about the fiber inclination $\alpha$.

The setup for the XP measurement is similar to the 3D-PLI measurement, but it does not include the retarder (see \cref{fig:Setups}b). 
For this setup, the transmitted light intensity (cf.\ \cref{fig:Setups}e) and the corresponding Fourier coefficients read:
\begin{align}
\vec{S}'_X(\rho) &= P_y\,(\rho,\,D_y,\,\tau_y) \cdot M\,(\varphi,\,\delta,\,D,\,\tau) \cdot P_x\,(\rho,\,D_x,\,\tau_x) \cdot \vec{S}_{\text{unpol}} \\ 
\begin{split}
\Rightarrow \,\,\, I_X(\rho) &= \tau \, \tau_x \, \tau_y \, I_0 \, 
\bigg[ 1 - D_x\,D_y
+ \Big(1 - \sqrt{1-D^2}\,\cos\delta \Big) \, \sin^2\big(2(\rho-\varphi)\big) \\
&\quad\quad\quad + D\,(D_x - D_y)\,\cos\big(2(\rho-\varphi)\big)
\bigg] 
\end{split} 
\\ \notag \\
\Rightarrow \,\,\, I_X(\rho) &= a_{0X} + a_{2X} \, \cos(2\rho) + b_{2X} \, \sin(2\rho) + a_{4X} \, \cos(4\rho) + b_{4X} \, \sin(4\rho), \\
a_{0X} &= \tau \, \tau_x \, \tau_y \, I_0 \, \left( 1 - \frac{1}{2} D_x \, D_y \, \left(1 + \sqrt{1-D^2}\,\cos\delta \right) \right), 
\label{eq:a0X}\\
a_{2X} &= \tau \, \tau_x \, \tau_y \, I_0 \, D \, \big(D_x - D_y \big) \, \cos(2\varphi), 
\label{eq:a2X}\\
b_{2X} &= \tau \, \tau_x \, \tau_y \, I_0 \, D \, \big(D_x - D_y \big) \, \sin(2\varphi), 
\label{eq:b2X}\\
a_{4X} &= - \frac{1}{2}\tau \, \tau_x \, \tau_y \, I_0 \, D_x \, D_y \, \left(1 - \sqrt{1-D^2}\,\cos\delta \right) \, \cos(4\varphi), 
\label{eq:a4X}\\
b_{4X} &= - \frac{1}{2}\tau \, \tau_x \, \tau_y \, I_0 \, D_x \, D_y \, \left(1 - \sqrt{1-D^2}\,\cos\delta \right) \, \sin(4\varphi) 
\label{eq:b4X}.
\end{align}

The direction angle of the fibers is given by the minima of the intensity signal (see \cref{fig:Setups}f, in red) and can be computed from the Fourier coefficients of fourth order via:\footnote{In principle, $\varphi_X$ could be computed from the Fourier coefficients of second order for a value range of $[0,\pi)$. However, this is not feasible because $a_{2X}$ and $b_{2X}$ are much smaller than $a_{4X}$ and $b_{4X}$ (the diattenuation of brain tissue is expected to be small and the linear polarizers have a similar degree of polarization $D_x \approx D_y$), resulting in a much lower signal-to-noise ratio.}
\begin{align}
\varphi_X &= \frac{\atantwo \big(b_{4X}, \, a_{4X}\big)}{4} \,\, \,\,
\in \,\, [0, \pi/2).
\label{eq:phi_X}
\end{align}


\subsection{Diattenuation Imaging (DI)}
\label{sec:Diattenuation_Measurement}

To measure the diattenuation $D$ of the brain tissue, only the polarizer is rotated below the stationary tissue sample (see \cref{fig:Setups}c) \cite{mehta2013}. For this setup, the transmitted light intensity (cf.\ \cref{fig:Setups}g) and the corresponding Fourier coefficients read:
\begin{align}
\vec{S}'_D(\rho) &= M\,(\varphi,\,\delta,\,D,\,\tau) \cdot P_x\,(\rho,\,D_x,\,\tau_x) \cdot \vec{S}_{\text{unpol}}  \\
\Rightarrow \,\,\, I_D(\rho) &= \tau \, \tau_x \, I_0 \, \Big( 1 + D\,D_x\,\cos\big(2(\rho-\varphi)\big) \Big) \\ \notag \\
\Rightarrow \,\,\, I_D(\rho) &= a_{0D} + a_{2D} \, \cos(2\rho) + b_{2D} \, \sin(2\rho), \\
a_{0D} &= \tau \, \tau_x \, I_0, 
\label{eq:a0D}\\
a_{2D} &= \tau \, \tau_x \, I_0 \,\,\, D \, D_x \, \cos(2\varphi), 
\label{eq:a2D}\\
b_{2D} &= \tau \, \tau_x \, I_0 \,\,\, D \, D_x \, \sin(2\varphi) 
\label{eq:b2D}.
\end{align}

The direction angle of the fibers is related to the rotation angle for which the transmitted light intensity $I_D(\rho)$ becomes maximal (see \cref{fig:Setups}g, in red) and can be computed from the Fourier coefficients of second order via:
\begin{align}
\Rightarrow \,\, \varphi_D &= \frac{\atantwo \big(b_{2D}, \, a_{2D}\big)}{2} \,\,\,\, \in \,\, [0, \pi).
\label{eq:phi_D}
\end{align}
As for the XP measurement, the determined direction angle does not depend on $D_x$, $D_y$, $\gamma$, or $D$.\footnote{However, the direction angle $\varphi_D$ obtained from the DI measurement is expected to be more error-prone than the direction angle $\varphi_X$ obtained from the XP measurement because $a_{2D}$ and $b_{2D}$ are much smaller than $a_{4X}$ and $b_{4X}$ leading to a smaller signal-to-noise ratio (the amplitude of the diattenuation signal $D$ is expected to be much smaller than the amplitude of the retardation signal $\sin^2(\delta/2)$, cf.\ \cref{fig:Setups}f and g).} 

The diattenuation of the brain tissue corresponds to the amplitude of the normalized intensity profile (see \cref{fig:Setups}f, in blue) and can be computed by combining all three Fourier coefficients:
\begin{align}
\Rightarrow \,\, D_D &= \frac{\sqrt{a^2_{2D} + b^2_{2D}}}{D_x \,\, a_{0D}}. 
\label{eq:D_Diat}
\end{align}


\section{Numerical Study}
\label{sec:Numerical_Study}

The analytical expressions in \cref{sec:Measurement_Setups_Analysis} were derived assuming an ideal light source and camera ($\vec{S}_L = \vec{S}_c = \vec{S}_{\text{unpol}}$). 
However, a thorough characterization of the optical system (see \cref{sec:Characterization_LAP}) has shown that the light emitted by the light source is slightly linearly polarized and that the camera is slightly sensitive to linearly and left-handed circularly polarized light. The study also revealed that the employed filters are not completely ideal: The polarizer and the analyzer have a degree of polarization slightly less than 100\,\%, i.\,e.\ the diattenuation of the polarizer ($D_x$) and the diattenuation of the analyzer ($D_y$) are less than one. Furthermore, the actual retardance $\gamma$ of the employed quarter-wave retarder differs from $\pi/2$ because the optimal working wavelength of the retarder does not perfectly match the illumination wavelength of the light source \cite{reckfort2015}. 
As the polarization effects are non-multiplicative and influence each other, the polarization properties of the optical components were determined as an average over the field of view. The diattenuation of the linear polarizers, the retardance of the quarter-wave retarder, and the normalized Stokes vectors of light source ($\vec{S}_L$) and camera ($\vec{S}_c$) were computed as (see \ref{sec:Characterization_Pol-Prop}):
\begin{align}
D_x \approx 0.98,  \,\,\,\,\,\,\,\,
D_y \approx 0.97,   \,\,\,\,\,\,\,\,
\gamma \approx 0.49\,\pi, \,\,\,\,\,\,\,\,
\vec{S}_L
\approx
\begin{pmatrix} 1 \\
				-5 \times 10^{-3} \\
				\,\,\,\,\,\, 8 \times 10^{-4} \\
				-5 \times 10^{-7}
\end{pmatrix}, \,\,\,\,\,\,\,\,
\vec{S}_c
\approx
\begin{pmatrix} 1 \\
				\,\,\,\,\,\, 8 \times 10^{-3} \\
				-1 \times 10^{-3} \\
				-5 \times 10^{-4}
\end{pmatrix}.
\label{eq:Polarization_Properties}
\end{align}

As the diattenuation of brain tissue is expected to be small (cf.\ \cref{sec:Introduction}), the measured diattenuation signal might be influenced by these non-ideal system properties.
To predict the impact of the non-ideal system properties and the tissue diattenuation on the reconstructed fiber orientations and the measured diattenuation, a numerical study was performed prior to the experimental study in \cref{sec:Experimental_Study_on_Brain_Tissue}. By combining the extended analytical model from \cref{sec:Measurement_Setups_Analysis} and the polarization properties defined in \cref{eq:Polarization_Properties}, the errors on the measured fiber orientation and diattenuation were estimated for the 3D-PLI, XP, and DI measurements, assuming arbitrary fiber orientations and tissue diattenuations.


\subsection{Methods}
\label{sec:Numerical_Methods}

For each measurement setup (3D-PLI, XP, DI), the expected transmitted light intensities were computed numerically using the polarization parameters defined in \cref{eq:Polarization_Properties}. 
The light source and the camera were described by the determined Stokes vectors $\vec{S}_L$ and $\vec{S}_c$. 
The filters (polarizer, retarder, analyzer) were described by the Müller matrices ($P_x$, $\Lambda$, $P_y$) as defined in \cref{eq:Analyzer_non-ideal,eq:Retarder_non-ideal,eq:Polarizer_non-ideal} using the determined values for $D_x$, $D_y$, and $\gamma$. 
The brain tissue was described by the Müller matrix $M$ as defined in \cref{eq:BrainTissue_non-ideal} with variables $\varphi$, $\delta$, and $D$. For reasons of simplification, the average transmittance of each filter and the intensity of the light source were set to one ($\tau_x = \tau_y = \tau_{\Lambda} = \tau = I_0 = 1$).

To compute the transmitted light intensities for each type of measurement, the Stokes vectors were multiplied with the corresponding Müller matrices (as described in \cref{sec:3D-PLI,sec:Crossed_Polars_Measurement,sec:Diattenuation_Measurement}) and the first entry of the resulting Stokes vectors was evaluated, respectively. To account for the image calibration performed for the tissue measurements (see \cref{sec:Characterization_Conclusion}), the resulting intensity profiles of the 3D-PLI and DI measurements were divided by the intensity profiles obtained from a matrix multiplication without the tissue matrix $M$. The image calibration for the XP measurement uses transmittance images of the filters and the light source (see \cref{sec:Characterization_Pol-Ind-Inhom,sec:Characterization_Conclusion}) and is already taken into account by setting the average transmittances and the intensity of the light source to one.

The numerically computed intensity profiles were analyzed as described in \cref{sec:3D-PLI,sec:Crossed_Polars_Measurement,sec:Diattenuation_Measurement}, respectively, and the direction angles ($\varphi_P$, $\varphi_X$, $\varphi_D$), the inclination angle ($\alpha_P$), and the diattenuation ($D_D$) were calculated from the determined Fourier coefficients.
As $D_x$ cannot be determined pixel-wise, $D_D$ and $D_x$ cannot be separated in a DI measurement. Therefore, the numerical and experimental studies investigate the amplitude of the diattenuation signal $\mathscr{D} \equiv D_D \, D_x$. To avoid confusion with the tissue diattenuation $D$, the symbol $\mathscr{D}$ will be referred to as \textit{measured diattenuation}.
The inclination angle $\alpha_P$ was computed from $\delta_P$ assuming that horizontal fibers (with $\alpha=0^{\circ}$) act as an ideal quarter-wave retarder, i.\,e.\ $\delta_P = (\pi/2)\,\cos^2\alpha_P$ (cf.\ \cref{eq:delta_approx}). To account for the correction with the maximum retardation value, the modified inclination angle $\tilde{\alpha}_P$ was computed using \cref{eq:alpha_corr} and $r_{\text{max}} = \pi/2$.

The impact of the non-ideal polarization properties was estimated by comparing the derived parameters (fiber orientation and diattenuation) to the tissue variables $\varphi$, $\alpha$, and $D$.
To enable a comparison with the experimental study in \cref{sec:Experimental_Study_on_Brain_Tissue}, a special focus was placed on the range $D \leq 4\,\%$. 


\subsection{Results}
\label{sec:Numerical_Results}

\subsubsection{Simulation of the 3D-PLI measurement}
\label{sec:3DPLI_Numerical}

\begin{figure}[h]
\centering
\includegraphics[width=1 \textwidth]{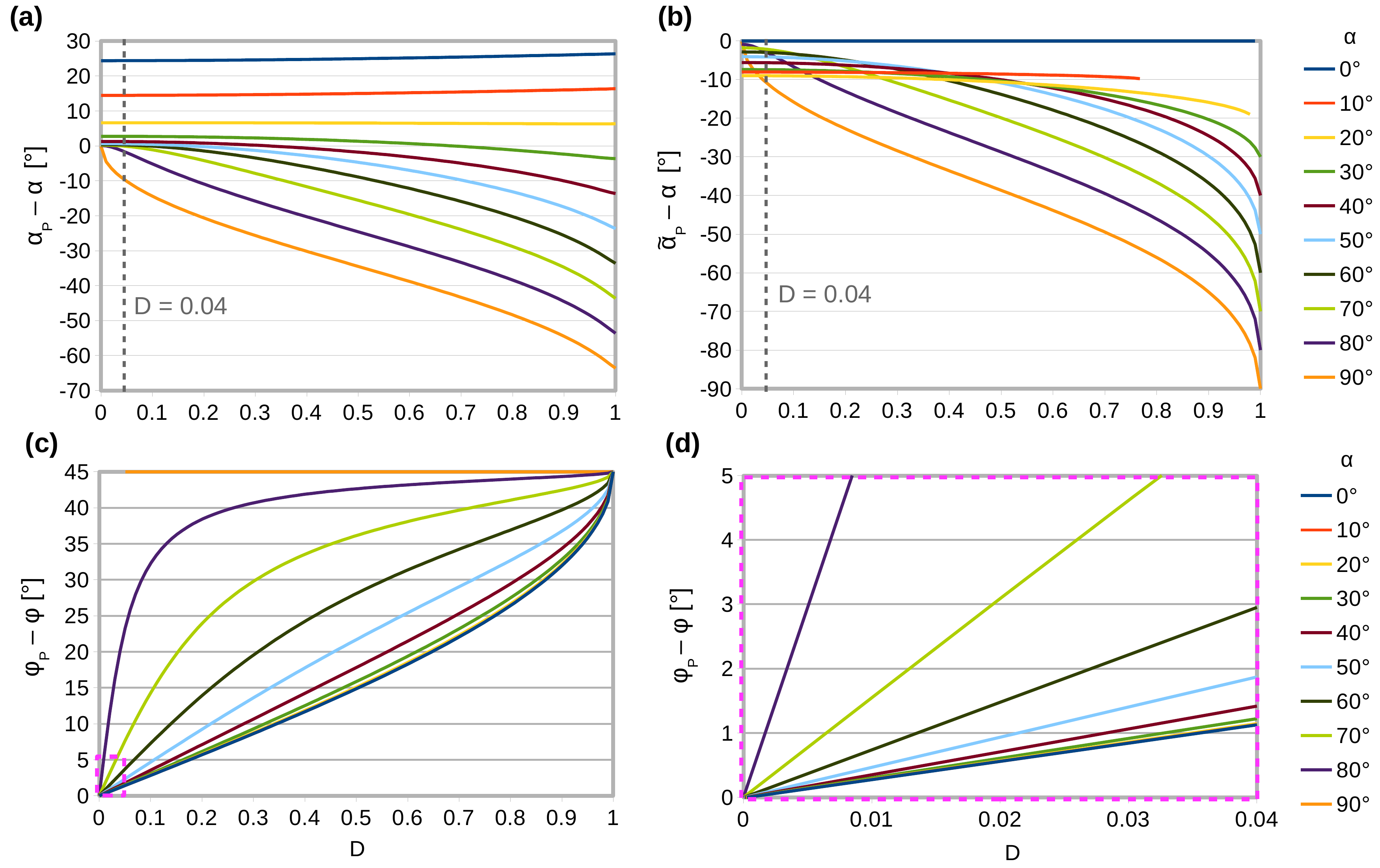}
\caption{Predicted impact of the non-ideal system properties (specified in \cref{eq:Polarization_Properties}) on the fiber orientation ($\varphi_P$, $\alpha_P$) measured from the simulation of the 3D-PLI measurement (see \cref{sec:3D-PLI}). The plots show the difference between ($\varphi_P$, $\alpha_P$) and the actual fiber orientation ($\varphi$, $\alpha$) for different $\alpha$ and tissue diattenuations $D$. The fiber direction is assumed to be $\varphi = 0^{\circ}$: \textbf{(a)} Difference between $\alpha_P$ and $\alpha$. \textbf{(b)} Difference between $\tilde{\alpha}_P$ and $\alpha$. \textbf{(c)} Difference between $\varphi_P$ and $\varphi$. \textbf{(d)} Enlarged view of (c) for $D \leq 4\,\%$.}
\label{fig:Numerical_3D-PLI}
\end{figure}

Figure \ref{fig:Numerical_3D-PLI} shows the predicted impact of the non-ideal system properties (specified in \cref{eq:Polarization_Properties}) on the measured inclination angles $\alpha_P$, $\tilde{\alpha}_P$, and the direction angle $\varphi_P$ for different tissue diattenuations $D$ and fiber inclinations $\alpha$. As the curves for different $\varphi$ look identical, the curves are only shown for an assumed fiber direction of $\varphi = 0^{\circ}$.

Figure \ref{fig:Numerical_3D-PLI}a shows that for fibers with smaller inclination angles ($\alpha < 30^{\circ}$), the measured inclination angle $\alpha_P$ is over-estimated ($\alpha_P > \alpha$) for all tissue diattenuations.
For fibers with larger inclination angles ($\alpha > 30^{\circ}$), the measured inclination angle is under-estimated ($\alpha_P < \alpha$) and the predicted difference between $\alpha$ and $\alpha_P$ increases with increasing tissue diattenuation. 
For $D \leq 4\,\%$, the maximum deviation from the actual fiber inclination is predicted to be about $25^{\circ}$ for fibers with $\alpha = 0^{\circ}$. 
After correcting with the maximum retardation value, the measured inclination angle $\tilde{\alpha}_P$ is under-estimated ($\tilde{\alpha}_P < \alpha$) for all inclination angles (see \cref{fig:Numerical_3D-PLI}b). The difference between $\alpha$ and $\tilde{\alpha}_P$ increases with increasing tissue diattenuation. For $D \leq 4\,\%$, the predicted difference is less than $10^{\circ}$ for all inclinations $< 90^{\circ}$.

Figure \ref{fig:Numerical_3D-PLI}c shows the influence of the non-ideal system components and tissue diattenuation on the measured direction angle $\varphi_P$. For all $D > 0$, the measured direction angle is expected to be over-estimated ($\varphi_P > \varphi$). The difference between $\varphi_P$ and $\varphi$ increases with increasing tissue diattenuation and fiber inclination. The maximum difference is $45^{\circ}$ for $D=1$ or $\alpha=90^{\circ}$. For small diattenuations ($D \leq 4\,\%$), the difference between $\varphi_P$ and $\varphi$ increases linearly with the tissue diattenuation (see \cref{fig:Numerical_3D-PLI}d). For fibers with $\alpha \leq 60^{\circ}$, the maximum difference is less than $3^{\circ}$. For steeper fibers, it is much larger.


\subsubsection{Simulation of the XP measurement}
\label{sec:CrossedPolars_Numerical}

\begin{figure}[htbp]
\centering
\includegraphics[width=0.9 \textwidth]{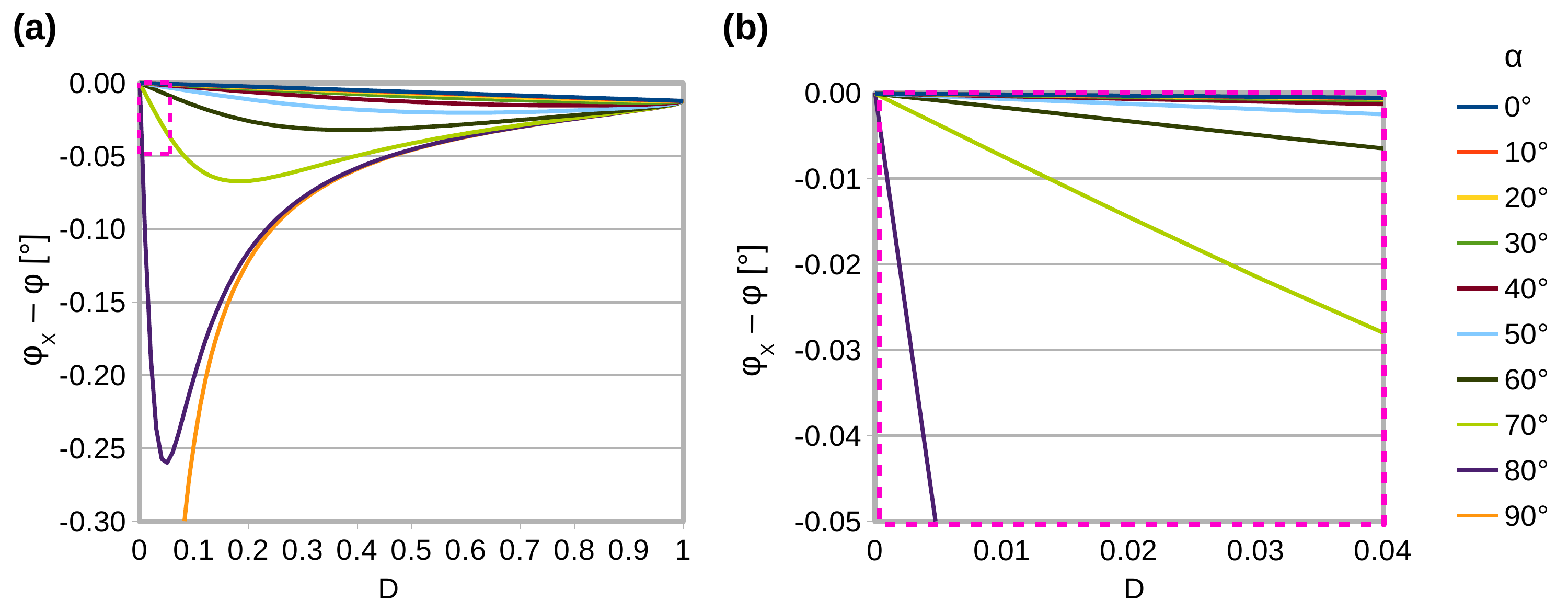}
\caption{Predicted impact of the non-ideal system properties (specified in \cref{eq:Polarization_Properties}) on the direction angle $\varphi_X$ derived from the simulation of the XP measurement (see \cref{sec:Crossed_Polars_Measurement}). The plots show the difference between $\varphi_X$ and the actual fiber direction $\varphi$ plotted against the tissue diattenuation $D$ for different fiber inclination angles $\alpha$. The actual fiber direction was set to $\varphi = 45^{\circ}$ such that $\vert \varphi_X - \varphi \vert$ becomes maximal. \textbf{(a)} Difference between $\varphi_X$ and $\varphi$. \textbf{(b)} Enlarged view of (a) for $D \leq 4\,\%$.}
\label{fig:Numerical-CrossedPolars}
\end{figure}

Figure \ref{fig:Numerical-CrossedPolars} shows the predicted impact of the non-ideal system properties (specified in \cref{eq:Polarization_Properties}) on the measured direction angle $\varphi_X$ for different tissue diattenuations $D$ and fiber inclinations $\alpha$. 

As can be seen in \cref{fig:Numerical-CrossedPolars}a, the direction angle $\varphi_X$ determined from the XP measurement is expected to deviate only slightly from the actual fiber direction $\varphi$ for all tissue diattenuations and inclinations $< 90^{\circ}$. For $\alpha \leq 70^{\circ}$, the difference is less than $0.08^{\circ}$. For $D \leq 4\,\%$, it is even less than $0.03^{\circ}$ (see \cref{fig:Numerical-CrossedPolars}b).


\subsubsection{Simulation of the DI measurement}
\label{sec:Diattenuation_Numerical}

\begin{figure}[htbp]
\centering
\includegraphics[width=1 \textwidth]{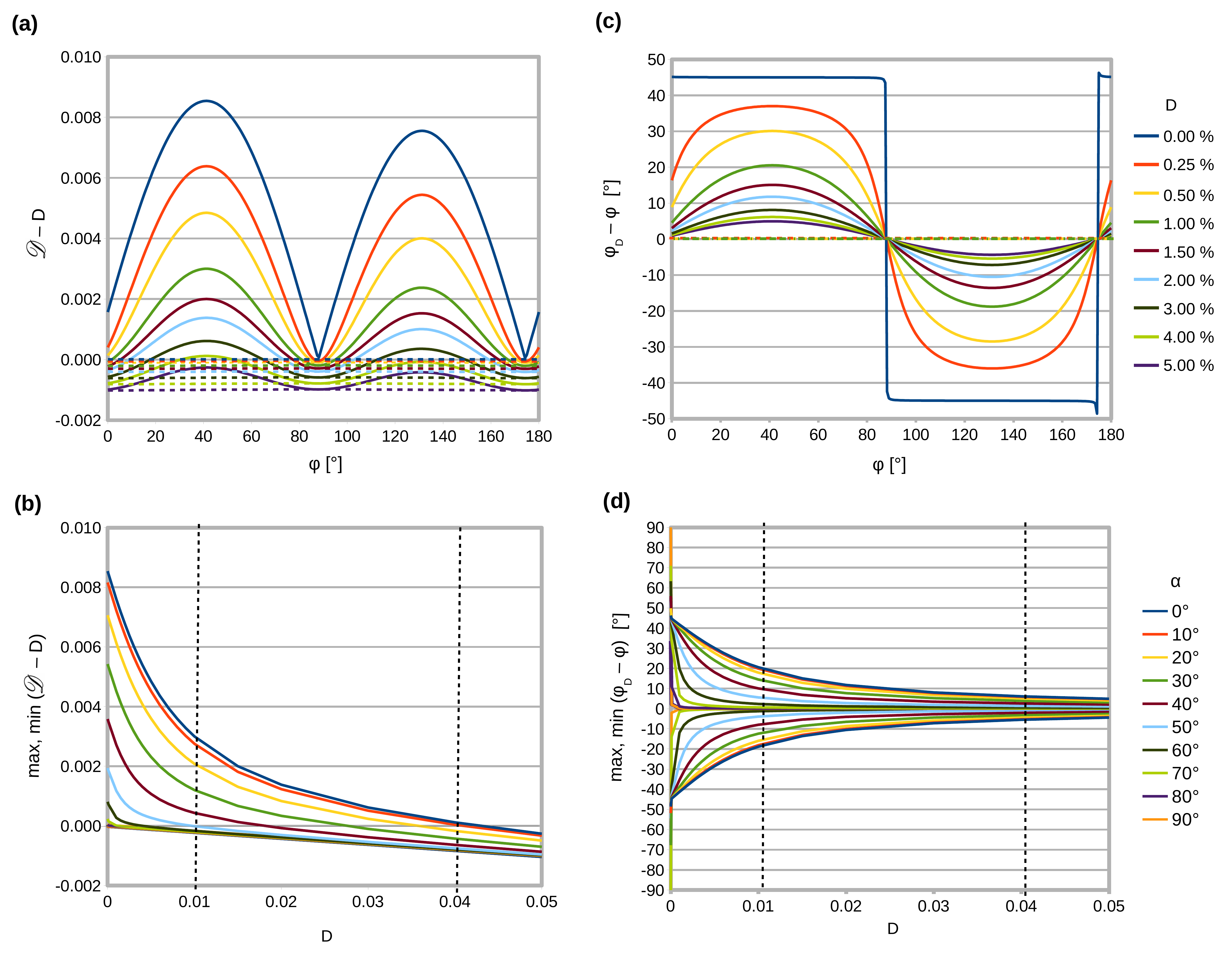}
\caption{Predicted impact of the non-ideal system properties (specified in \cref{eq:Polarization_Properties}) on the diattenuation $\mathscr{D}$ and the direction angle $\varphi_D$ derived from the simulation of the DI measurement (see \cref{sec:Diattenuation_Measurement}): \textbf{(a)} Difference between $\mathscr{D}$ and the actual tissue diattenuation $D$ plotted against the fiber direction $\varphi$ for different $D$. The solid lines correspond to horizontal fibers with inclination $\alpha = 0^{\circ}$, the dashed lines to vertical fibers with $\alpha = 90^{\circ}$. The curves for fibers with intermediate inclination angles lie in between. \textbf{(b)} Maximum difference (pos./neg.) between $\mathscr{D}$ and $D$ plotted against $D$ for different $\alpha$. Note that the values min($\mathscr{D} - D$) correspond to the bottom curve for all $\alpha$. \textbf{(c)} Difference between $\varphi_D$ and $\varphi$ plotted against $\varphi$ for different $D$. The solid lines correspond to horizontal fibers with inclination $\alpha = 0^{\circ}$, the dashed lines (all lying along the zero line) to vertical fibers with $\alpha = 90^{\circ}$ and $D > 0$. The curves for fibers with intermediate inclination angles lie in between. \textbf{(d)} Maximum difference (pos./neg.) between $\varphi_D$ and $\varphi$ plotted against $D$ for different $\alpha$.}
\label{fig:Numerical-Diattenuation}
\end{figure}

Figure \ref{fig:Numerical-Diattenuation} shows the predicted impact of the non-ideal system properties (specified in \cref{eq:Polarization_Properties}) on the measured diattenuation $\mathscr{D}$ and direction angle $\varphi_D$ for different tissue diattenuations $D$, fiber directions $\varphi$, and inclinations $\alpha$. 

As can be seen in \cref{fig:Numerical-Diattenuation}a, the measured diattenuation $\mathscr{D}$ depends on the fiber orientation ($\varphi$, $\alpha$) as well as on the diattenuation $D$ of the brain tissue. For regions with no tissue diattenuation ($D=0\,\%$), the measured diattenuation reaches values up to $\mathscr{D} = 0.85\,\%$ (for $\alpha=0^{\circ}$, $\varphi \approx 41^{\circ}$). 
The measured diattenuation is mostly over-estimated for $D<4\,\%$. For larger tissue diattenuations or steep fibers, $\mathscr{D}$ is slightly under-estimated. For $1\,\% \leq D \leq 4\,\%$, the difference between $\mathscr{D}$ and $D$ lies between $-0.1\,\%$ and $+0.3\,\%$ for all fiber inclinations (see \cref{fig:Numerical-Diattenuation}b).

The measured direction angle $\varphi_D$ corresponds only to the actual fiber direction for steep fibers, large tissue diattenuations, or $\varphi \approx 88^{\circ} (\pm 90^{\circ})$ (see \cref{fig:Numerical-Diattenuation}c). The maximum absolute difference between measured and actual fiber direction angle decreases with increasing tissue diattenuation (see \cref{fig:Numerical-Diattenuation}d): For regions with $D=1\,\%$, the maximum absolute difference is about $21^{\circ}$ (for $\alpha=0^{\circ}$, $\varphi \approx 41^{\circ}$). For regions with $D=4\,\%$, it is only about $6^{\circ}$.


\subsection{Discussion}
\label{sec:Numerical_Discussion}

The numerical study investigated the influence of the non-ideal system properties and the tissue diattenuation on the measured fiber orientation and diattenuation, allowing to make predictions and error estimations for the experimental study in \cref{sec:Experimental_Study_on_Brain_Tissue}.

When interpreting the numerical results, we should keep in mind that for perfectly vertical fibers, the direction angle is not defined and the tissue diattenuation is expected to be zero (if the vertical fibers are radially symmetric). The simulated graphs for $\alpha=90^{\circ}$ are therefore limiting cases and were only shown for reasons of completeness. For steep fibers, large deviations of the measured direction angle have only a very small influence on the overall fiber orientation vector and the tissue diattenuation is expected to be very small. Therefore, the simulated graphs for steep fibers should only be considered for small diattenuation values. 

The numerical study has shown that the correction with the maximum retardation value significantly improves the determination of the inclination angle $\alpha_P$ and should therefore be included in the 3D-PLI analysis. Before the correction, $\alpha_P$ is over-estimated up to $25^{\circ}$, afterwards it is only under-estimated up to $10^{\circ}$ for $D \leq 4\,\%$ and $\alpha < 90^{\circ}$.

The numerical study has also shown that the direction angle $\varphi_X$ is a good reference value for the actual fiber direction $\varphi$ as it is nearly independent from the diattenuation of the brain tissue and from the polarization properties of the optical components. 
In contrast to $\varphi_X$, the direction angles derived from the 3D-PLI and DI measurements ($\varphi_P$ and $\varphi_D$) depend on the polarization properties and on the tissue diattenuation.\footnote{Note that the sign of $(\varphi_P - \varphi)$ and $(\varphi_D - \varphi)$ will be flipped if the axis of maximum transmittance is orthogonal (and not parallel) to the fiber axis, which causes the graphs in \cref{fig:Numerical_3D-PLI}c,d and \cref{fig:Numerical-Diattenuation}c,d to be mirrored along the x-axis.}

For $D \leq 4\,\%$, the difference between $\varphi_P$ and $\varphi$ increases linearly with $D$ (see \cref{fig:Numerical_3D-PLI}d). However, the impact of the diattenuation is expected to be negligible in the 3D-PLI signal analysis: For $\alpha \leq 60^{\circ}$, the predicted difference between $\varphi_P$ and $\varphi$ is less than $3^{\circ}$. For steep fibers, we expect smaller diattenuation values, resulting in even smaller differences. 

While the difference between $\varphi_P$ and $\varphi$ is mostly independent from $\varphi$ and relatively small, the difference between $\varphi_D$ and $\varphi$ strongly depends on the actual fiber direction angle and is largest for small diattenuations and flat fibers (see \cref{fig:Numerical-Diattenuation}c). This shows that $\varphi_D$ is strongly influenced by the partially polarized light source and the polarization sensitivity of the camera.
The direction angle $\varphi_D$ derived from the DI measurement is broadly distributed around the actual fiber direction angle. However, the maximum absolute difference between $\varphi_D$ and $\varphi$ decreases with increasing $D$ (for $D=4\,\%$, it approaches $6^{\circ}$).

The measured diattenuation $\mathscr{D}$ is also strongly influenced by the non-ideal polarization properties of the light source and the camera. 
Without tissue diattenuation ($D=0$), the measured diattenuation reaches values up to $\mathscr{D} = 0.85\,\%$ (see \cref{fig:Numerical-Diattenuation}a). To ensure that the measured diattenuation mostly corresponds to the actual tissue diattenuation, the experimental study in \cref{sec:Experimental_Study_on_Brain_Tissue} focuses on values $\mathscr{D} > 1\,\%$. In this regime, the difference between $\mathscr{D}$ and $D$ is expected to be less than $0.3\,\%$ for $D \leq 4\,\%$ (see \cref{fig:Numerical-Diattenuation}b). 

As the polarization properties of light source, camera, and filters could only be determined as an average over the field of view, local deviations of $D_x$, $D_y$, $\gamma$, $\vec{S}_L$, and $\vec{S}_c$ were not included in the numerical study.
Thus, when comparing the predictions of the numerical study to the results of the experimental study, the graphs should only be considered as a reference.


\section{Experimental Study on Brain Tissue}
\label{sec:Experimental_Study_on_Brain_Tissue}

To quantify the diattenuation of brain tissue and its impact on the measured 3D-PLI signal, an experimental study was performed. In addition to the strength of the diattenuation signal, the correlation between the birefringence and the diattenuation of brain tissue was studied. Furthermore, it was analyzed how large the impact of the tissue diattenuation is on the fiber directions obtained from the DI and 3D-PLI measurements and whether the measured diattenuation signal contains structural information about the brain tissue. The experimental study was conducted exemplary on five sagittal sections of a healthy Wistar rat brain. 


\subsection{Methods}
\label{sec:Methods}

\subsubsection{Tissue preparation}
\label{sec:Tissue_Preparation}

\begin{figure}[htbp]
\centering
\includegraphics[width=0.9 \textwidth]{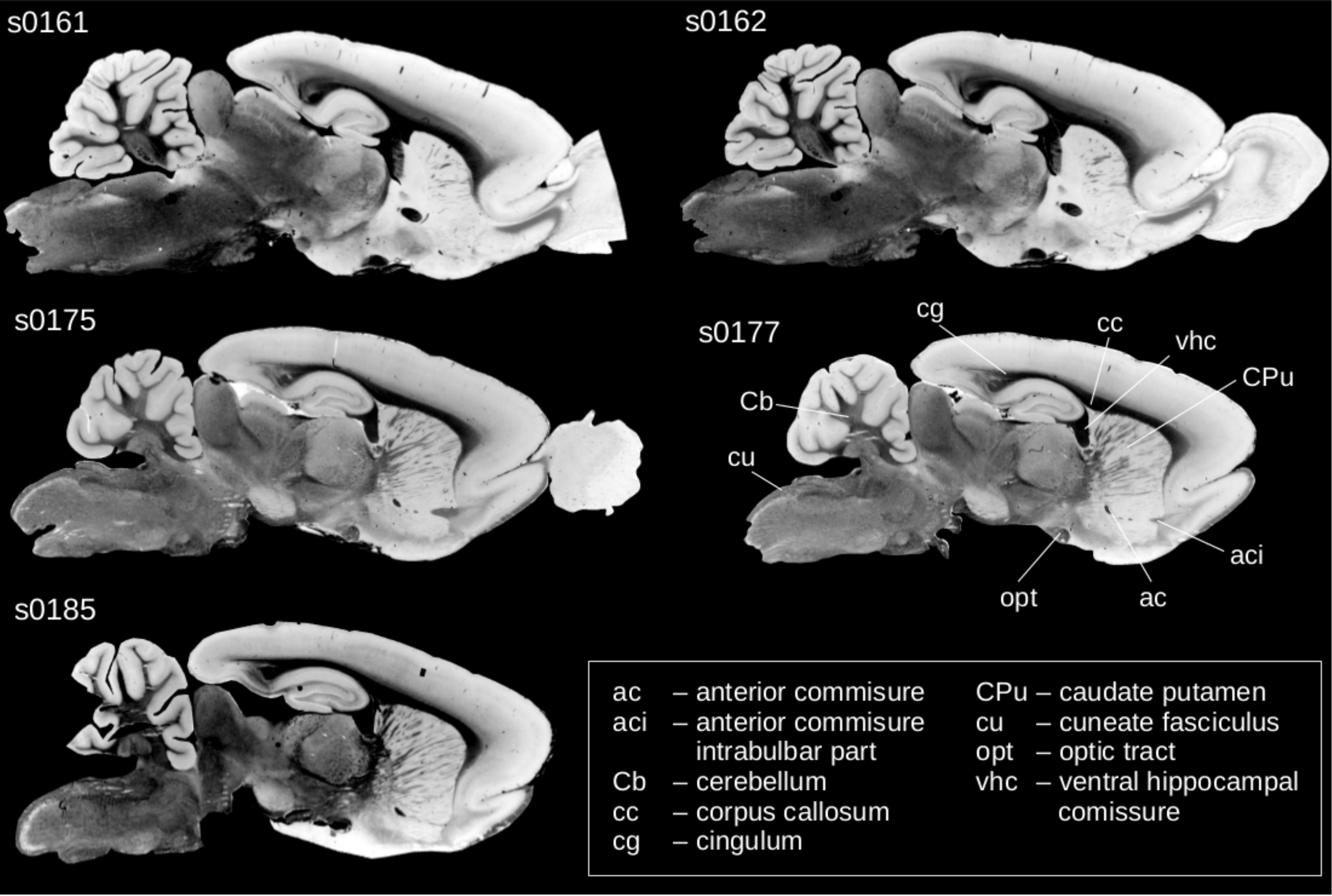}
\caption{Transmittance images of the five investigated rat brain sections. For reference, some anatomical structures are labeled exemplary in section s0177.}
\label{fig:Transmittance_ROIs}
\end{figure}

All animal procedures were approved by the institutional animal welfare committee at the Research Centre Jülich and were in accordance with European Union (National Institutes of Health) guidelines for the use and care of laboratory animals. 
Immediately after death, the rat brain was removed from the scull, fixed with $4\,\%$ buffered formaldehyde, immersed in a $20\,\%$ glycerin solution with Dimethyl sulfoxide for cryo-protection, and deep-frozen at $-80^{\circ}$C. The frozen brain was then cut along the sagittal plane with a cryotome 
into histological sections of 60\,\textmu m thickness. Five sections from the middle part (section numbers s0161, s0162, s0175, s0177, and s0185) were selected for evaluation (see \cref{fig:Transmittance_ROIs}). The brain sections were mounted on glass slides, embedded in a $20\,\%$ glycerin solution, cover-slipped, and measured 1-2 days afterwards.


\subsubsection{Polarimetric measurements}
\label{sec:Polarimetric_Measurements}

The 3D-PLI, XP, and DI measurements were performed with the polarimetric setups shown in \cref{fig:Setups}a-c, using an object space resolution of 14 \textmu m/px and an illumination wavelength of $525$\,nm. The optical properties of the polarimeter are specified in \cref{sec:Components_LAP}. In all cases, the filters were rotated around the stationary brain section in equidistant angles ($\rho = \{0^{\circ}, 10^{\circ}, \dots, 170^{\circ}\}$). For the 3D-PLI measurement, an image of the brain section was captured once for each rotation angle. As the signal-to-noise ratio of the XP and DI measurements is lower than for the 3D-PLI measurement (cf.\ \cref{fig:parametermaps} in \cref{sec:ParameterMaps}), the image of the brain section was recorded several times before rotating the filters (polarizer and/or analyzer) to the next angle position (10 times for the XP and 20 times for the DI measurement). The resulting images were averaged for each rotation angle. 

The calibration of the 3D-PLI, XP, and DI measurements were executed as proposed in \cref{sec:Characterization_Conclusion}. For the 3D-PLI measurement, a set of $50$ calibration images (without specimen) was taken for each of the $18$ rotation angles and an average image was computed for each rotation angle as proposed by Dammers et al.\ \cite{dammers2011}. The measured images of the specimen were divided by the corresponding averaged calibration image for each rotation angle and multiplied by the mode intensity of all $900$ calibration images. 
The images obtained from the DI measurement were calibrated in an equivalent way. For both calibrations, it has been ensured that the intensity of the light source did not change between calibration and specimen measurements. 
For the XP measurement, the parallax effect induced by the analyzer was corrected and the images were calibrated as described in \cref{sec:Characterization_Conclusion}. After calibration, all images show a uniform background intensity.


\subsubsection{Registration of images}
\label{sec:Registration_of_Images}

The different types of filters used for the 3D-PLI, XP, and DI measurements introduce different parallax effects to the measured images. Furthermore, the camera position might slightly shift when changing the measurement setup and the brain sections might slightly move over time due to the embedding in glycerin solution.
To still enable a pixel-wise comparison between the various measurements, the images were registered onto each other using the open source software \textit{elastix} which is based on a multi-resolution approach with four resolution levels \cite{elastix}. The registration was chosen to be linear affine and mutual information was employed as a metric. 
The amplitude of the diattenuation signal is much smaller than the amplitude of the birefringence signal measured in the 3D-PLI and XP measurements (cf.\ \cref{fig:parametermaps} in \cref{sec:ParameterMaps}). To minimize interpolation artifacts in the diattenuation images, the calibrated images of the 3D-PLI and XP measurements were registered onto the calibrated images of the DI measurement.


\subsubsection{Signal analysis}
\label{sec:Exp_Signal_Analysis}

The calibrated and registered image series were Fourier analyzed and the parameter maps (transmittance, direction, retardation, and diattenuation) were computed from the Fourier coefficients as described in \cref{sec:3D-PLI,sec:Crossed_Polars_Measurement,sec:Diattenuation_Measurement}. The parameter maps are shown exemplary for one rat brain section in \cref{sec:ParameterMaps}.

The transmittance $I_{T,P}$, the direction angle $\varphi_P$, and the retardation $r_P$ were computed from the average, the phase, and the amplitude of the measured 3D-PLI signal according to \cref{eq:transmittance,eq:direction,eq:retardation}. The direction angle $\varphi_X$ obtained from the XP measurement was computed using \cref{eq:phi_X}. The diattenuation $\mathscr{D}$ ($= D_D\,D_x$) and the direction angle $\varphi_D$ were determined from the amplitude and the position of the maximum transmitted light intensity of the measured DI signal according to \cref{eq:D_Diat,eq:phi_D}.

The signal analysis was performed for the complete brain sections. To ensure that the diattenuation signal is mainly caused by the brain tissue and not by non-ideal system components (cf.\ \cref{sec:Diattenuation_Numerical}), a special focus was placed on regions with $\mathscr{D}$ $> 1\,\%$.

The direction angle $\varphi_X$ was used as a reference for the actual fiber direction angle $\varphi$ (cf.\ \cref{sec:CrossedPolars_Numerical}). As mentioned in \cref{sec:Crossed_Polars_Measurement}, $\varphi_X$ can only be determined in a value range of $[0^{\circ},90^{\circ})$, while $\varphi_P$ and $\varphi_D$ have value ranges of $[0^{\circ},180^{\circ})$. When being compared to $\varphi_X$, the direction angles $\varphi_P$ and $\varphi_D$ were therefore reduced to a value range of $[0^{\circ},90^{\circ})$. In regions with small retardation values, the transmitted light intensity in the XP measurement is small due to the $90^{\circ}$-orientation of the linear polarizers  (cf.\ \cref{fig:Setups}f), leading to a low signal-to-noise ratio. Therefore, $\varphi_X$ was only evaluated in regions with retardation values $r_P > 0.1$ (cf.\ \cref{fig:DeltaDIR-XPvsRET}b in \cref{sec:DeltaDIR_vs_RET}).


\subsection{Results}
\label{sec:Results}

\subsubsection{Strength of the diattenuation signal}
\label{sec:Strength_of_diattenuation_signal}

The five investigated rat brain sections accumulate in total to over 4.000.000 investigated pixels. The selected regions with $\mathscr{D}>1\,\%$ represent about $6\,\%$ of the tissue. Considering all investigated pixels, the average diattenuation is about $0.42\,\%$ (with $0.34\,\%$ standard deviation). Taking only regions with $\mathscr{D}>1\,\%$ into account, the average is about $1.42\,\%$ (with $0.43\,\%$ standard deviation). The maximum measurable diattenuation signal ($\mathscr{D}_{\text{max}} \approx 3.9\,\%$) was observed within the optic tract.


\subsubsection{Correlation between the measured retardation and diattenuation}
\label{sec:Correlation_between_diattenuation_retardation}

\begin{figure}[htbp]
	\centering 
	\begin{tabular}{ll}
	\textbf{(a)}
	\begin{adjustbox}{valign=t}
	\begin{tabular}{@{}c@{}}
	\includegraphics[width=0.45 \textwidth]{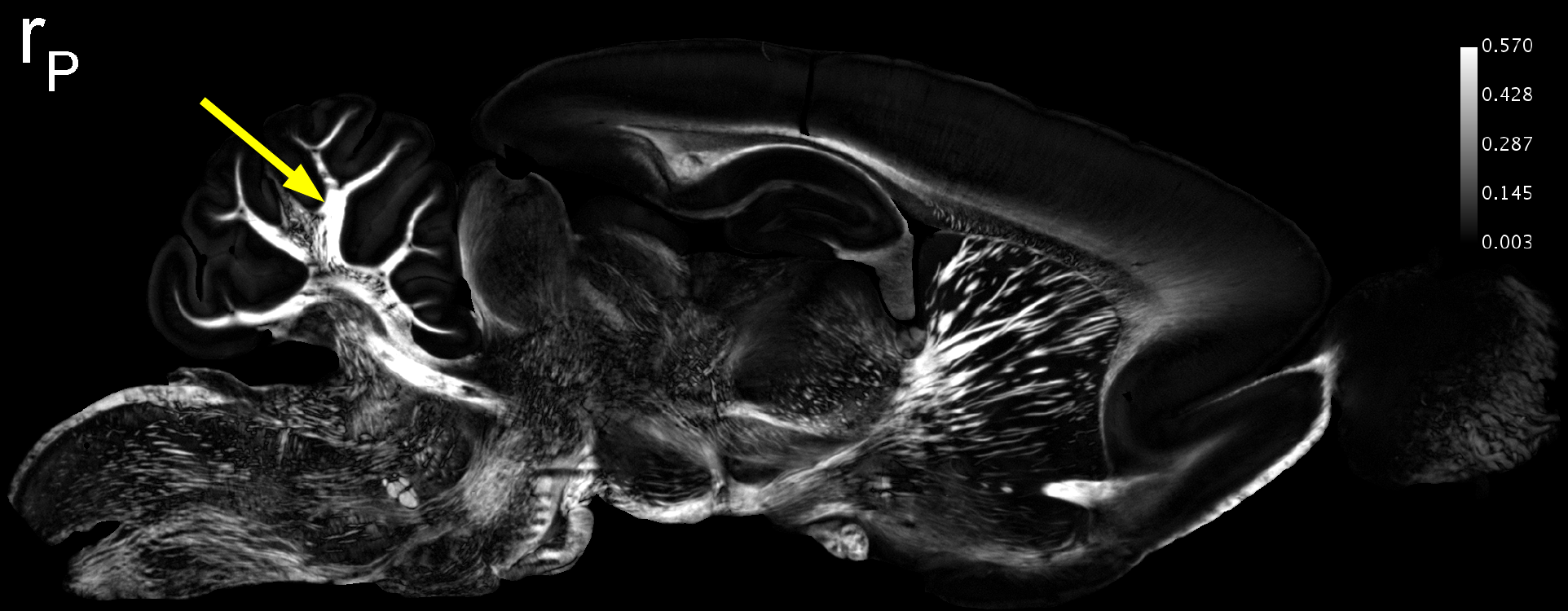} \\
  	\includegraphics[width=0.45 \textwidth]{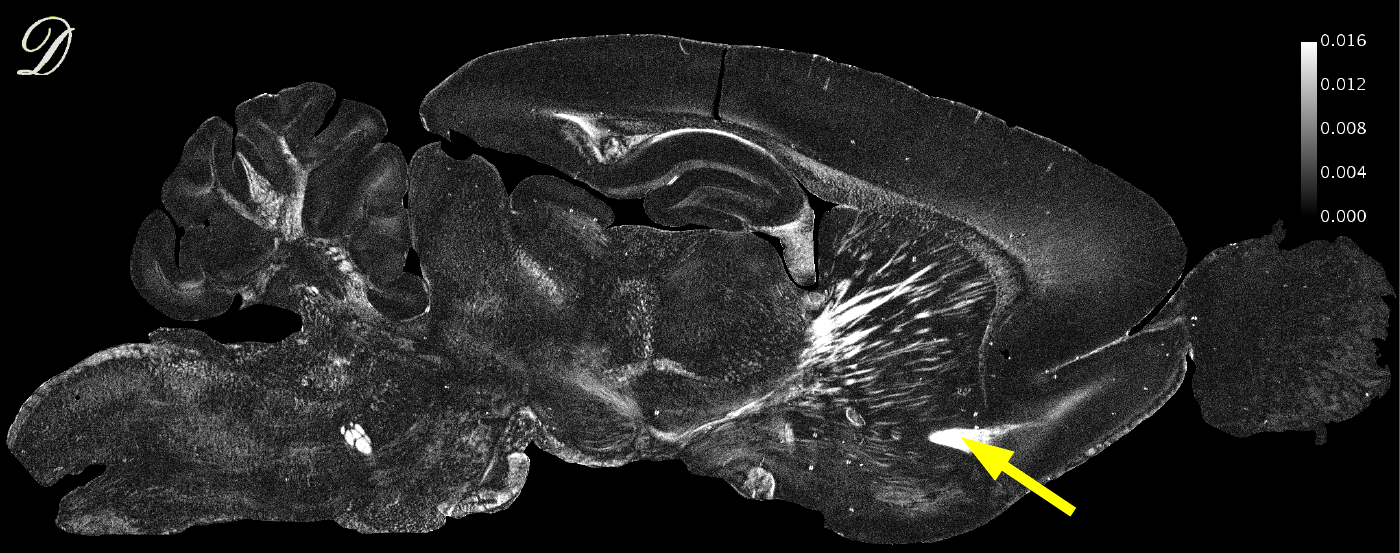}
	\end{tabular}
	\end{adjustbox}
	\textbf{\ \ (b)} 
	\begin{adjustbox}{valign=t}
	\includegraphics[width=0.45 \textwidth]{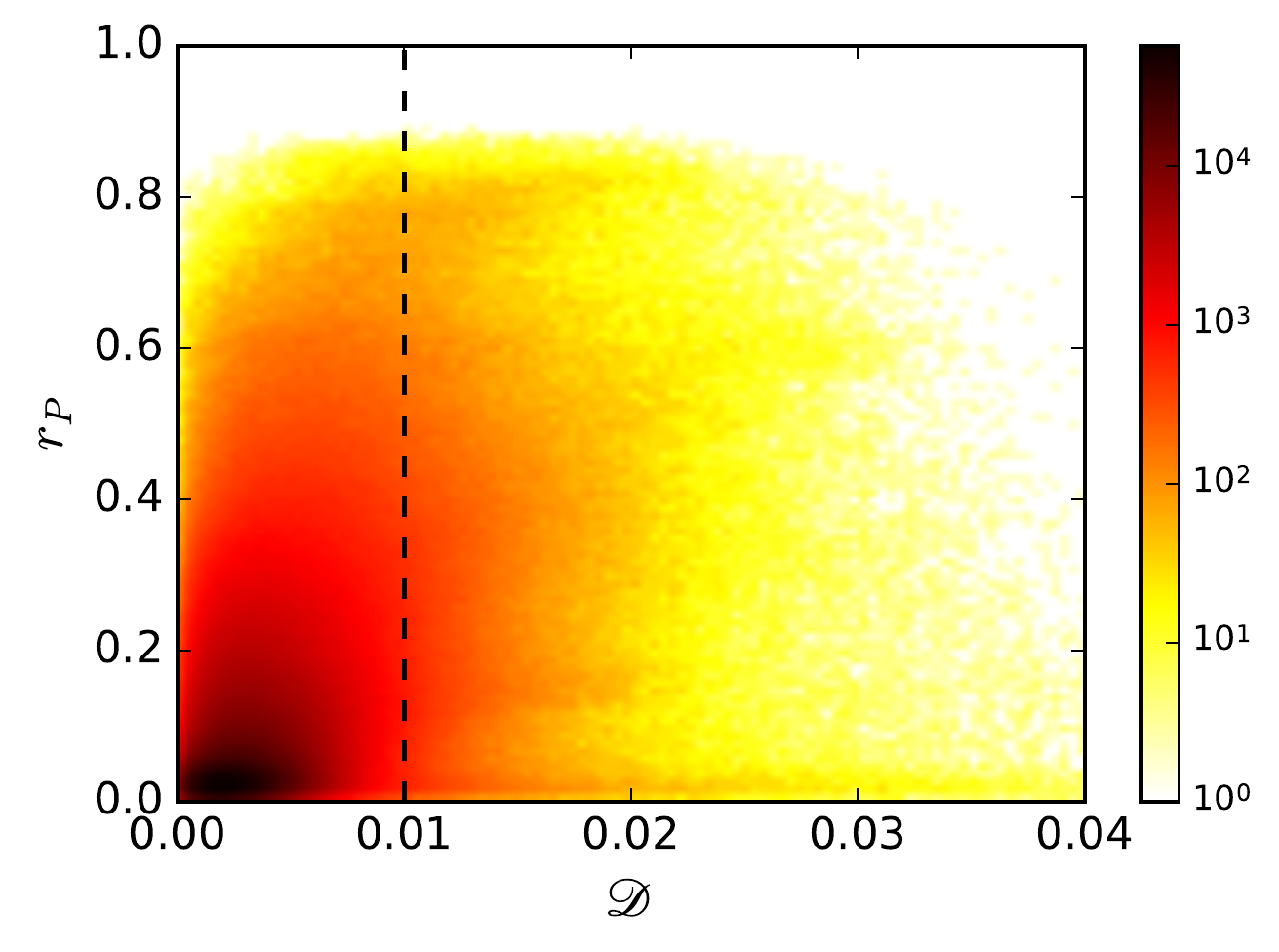}
	\end{adjustbox}
	\end{tabular}	
\caption{Correlation between measured retardation and diattenuation: \textbf{(a)} Retardation $r_P$ and diattenuation $\mathscr{D}$ shown exemplary for one rat brain section (s0175). The yellow arrows indicate the regions with maximum retardation and diattenuation, respectively. \textbf{(b)} 2D histogram showing $r_P$ plotted against $\mathscr{D}$ evaluated for all five brain sections. The number of bins is 100 for both axes. The dashed vertical line marks the region ($\mathscr{D}>1\,\%$) for which the diattenuation signal is expected to be mainly caused by the brain tissue and not by non-ideal system components (cf.\ \cref{sec:Diattenuation_Numerical}).}
\label{fig:DIAvsRET}
\end{figure}

Figure \ref{fig:DIAvsRET}a shows the measured retardation and diattenuation exemplary for one rat brain section (s0175). A qualitative comparison reveals that some regions with a high retardation signal (e.\,g.\ the anterior commissure intrabulbar part and the caudate putamen) also show a relatively high diattenuation signal and that some areas with a low retardation signal (e.\,g.\ gray matter regions) show a low diattenuation signal. These observations are consistent across the different brain sections. A quantitative comparison, however, reveals that the retardation signal does not necessarily correlate with the diattenuation signal: In the displayed brain section, the regions with maximum retardation and diattenuation are marked with a yellow arrow, respectively. While the largest diattenuation signal was measured within the anterior commissure intrabulbar part ($\overline{\mathscr{D}}(\text{aci})=2.1\,\%$, $\mathscr{D}_{\text{max}}(\text{aci})=3.5\,\%$), the largest retardation signal was measured within the cerebellum ($r_{P,\text{max}}(\text{cb})=0.78$) and not within the anterior commissure intrabulbar part ($r_{P,\text{max}}(\text{aci})=0.63$).

The 2D histogram in \cref{fig:DIAvsRET}b further manifests that there exists no distinct correlation between the strength of the retardation and the diattenuation. For $\mathscr{D} > 1\,\%$, regions with the same retardation value show very different diattenuation values and vice-versa (the correlation coefficient was determined to be about 0.1).


\subsubsection{Comparison of the fiber directions obtained from the DI and 3D-PLI measurements}
\label{sec:Fiber_direction_from_diattenuation}

\begin{figure}[htbp]
\centering
\includegraphics[width=0.5 \textwidth]{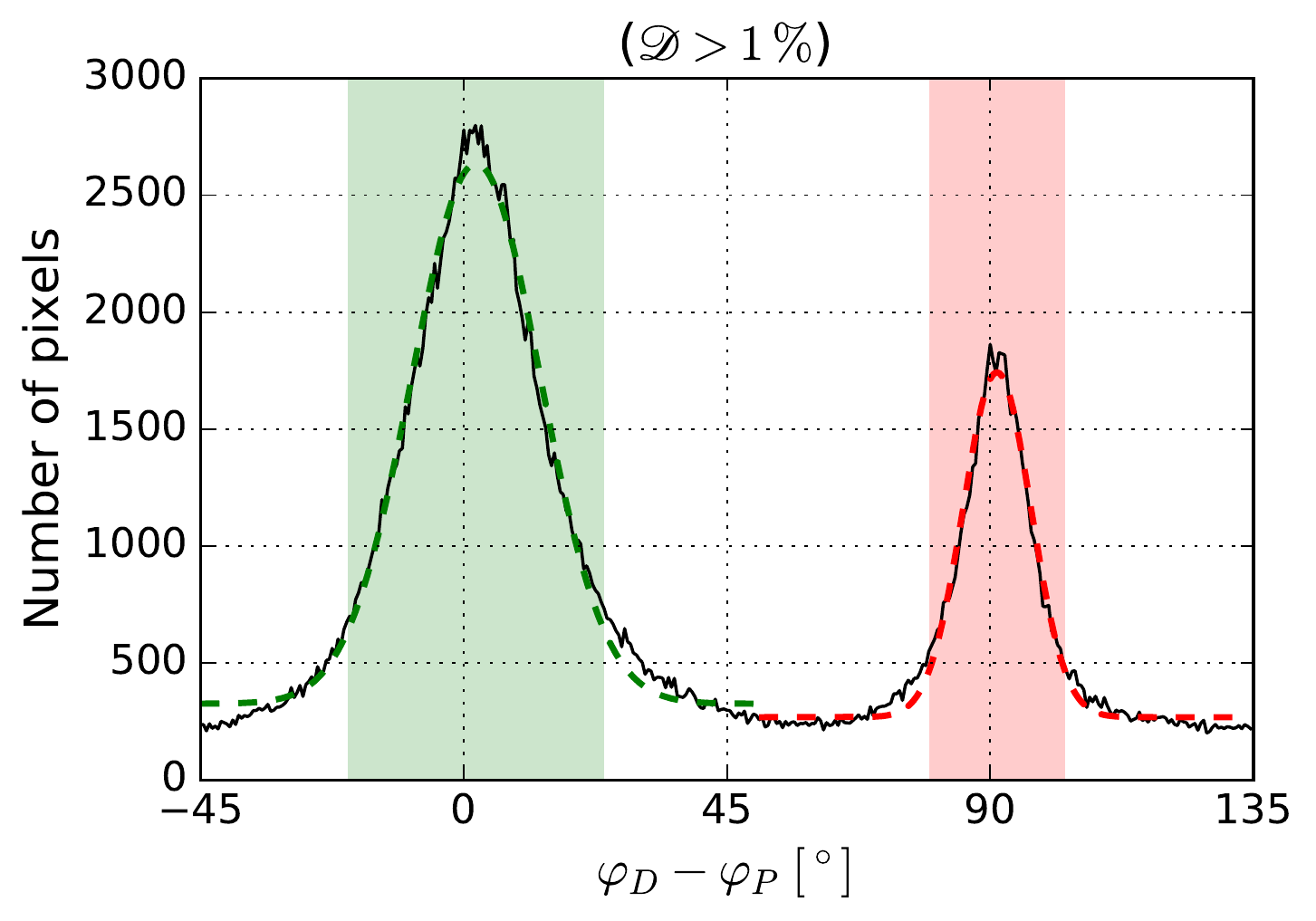}
\caption{Histogram showing the difference between the direction angle $\varphi_D$ determined from the DI measurement and the direction angle $\varphi_P$ determined from the 3D-PLI measurement (bin width $= 0.5^{\circ}$). Due to the $180^{\circ}$-periodicity, the data range has been reduced to $[-45^{\circ},135^{\circ})$. To ensure that the diattenuation signal is mainly caused by the brain tissue and not by non-ideal system components (cf.\ \cref{sec:Diattenuation_Numerical}), only regions with $\mathscr{D}>1\,\%$ were used for evaluation. The highlighted regions show the 2$\sigma$-environments of the fitted Gaussian distributions: green ($53.97\,\%$ of the selected pixels: $\mu = 2.1^\circ, \sigma=11.0^\circ$), red ($19.06\,\%$ of the selected pixels: $\mu = 91.1^\circ, \sigma=5.8^\circ$).}
\label{fig:Histo_DeltPhi_D-PLI}
\end{figure}

A comparison of the direction angle $\varphi_D$ determined from the DI measurement and the direction angle $\varphi_P$ determined from the 3D-PLI measurement reveals that the majority of the determined fiber directions is in good correspondence. The histogram of ($\varphi_D - \varphi_P$) in \cref{fig:Histo_DeltPhi_D-PLI} shows a distinct peak around $0^\circ$, highlighted in green. The peak can be described by a Gaussian distribution with mean $\mu = 2.1^\circ$ and standard deviation $\sigma = 11.0^\circ$. In certain brain regions, however, the fiber direction derived from the DI measurement is shifted by $90^\circ$, see peak highlighted in red. The peak can be described by a Gaussian distribution with $\mu = 91.1^\circ$ and $\sigma = 5.8^\circ$. 

For the derivation of $\varphi_D$ (see \cref{eq:phi_D}), it was assumed that the fiber direction corresponds to the rotation angle for which the transmitted light intensity becomes maximal. This assumption holds only for some brain regions (green area). In other brain regions (red area), the fiber direction is given by the rotation angle for which the transmitted light intensity becomes minimal, resulting in the observed $90^\circ$-shift.

For further investigation, all pixels belonging to the green highlighted area in \cref{fig:Histo_DeltPhi_D-PLI} ($\varphi_D - \varphi_P \in [-19.9^{\circ}, 24.1^{\circ}]$) will be denoted by $D^+$ and all pixels belonging to the red highlighted area ($\varphi_D - \varphi_P \in [79.5^{\circ}, 102.7^{\circ}]$) will be denoted by $D^-$. The angle ranges correspond to the $2\sigma$-environments of the Gaussian peaks (the other direction angles cannot be clearly assigned to $D^+$ or $D^-$).

\begin{figure}[htbp]
\centering
\includegraphics[width=0.9 \textwidth]{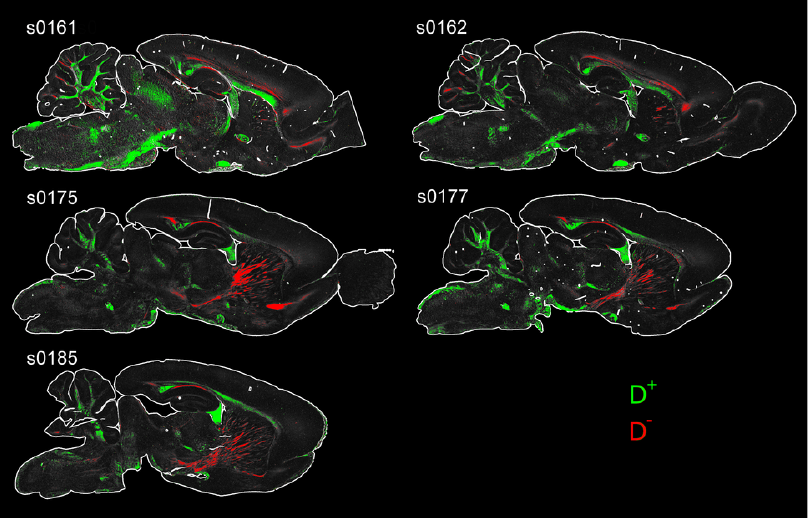}
\caption{Diattenuation images of the five investigated rat brain sections: Regions that show different types of diattenuation are highlighted in green ($D^{+}$) and red ($D^{-}$), corresponding to the angle ranges defined in \cref{fig:Histo_DeltPhi_D-PLI} (for $\mathscr{D}>1\,\%$).}
\label{fig:Maps_+-D}
\end{figure}

The type of diattenuation seems to be specific for certain brain regions. Some brain regions (ac, Cb, cu, opt, vhc, part of cc) show diattenuation of type $D^{+}$, while other brain regions (aci, CPu, part of cg) show diattenuation of type $D^{-}$ (see \cref{fig:Maps_+-D,fig:Transmittance_ROIs}). This behavior is consistent across the investigated brain sections.


\subsubsection{Impact of the tissue diattenuation on the fiber directions obtained from the DI and 3D-PLI measurements}
\label{sec:Impact_of_diattenuation_on_fiber_direction}

\begin{figure}[htbp]
\centering $
	\begin{array}{ll}
	\textbf{(a)} & \textbf{(b)} \\
	\includegraphics[height=0.43\textwidth, trim={0 0 1.85cm 0}, clip]{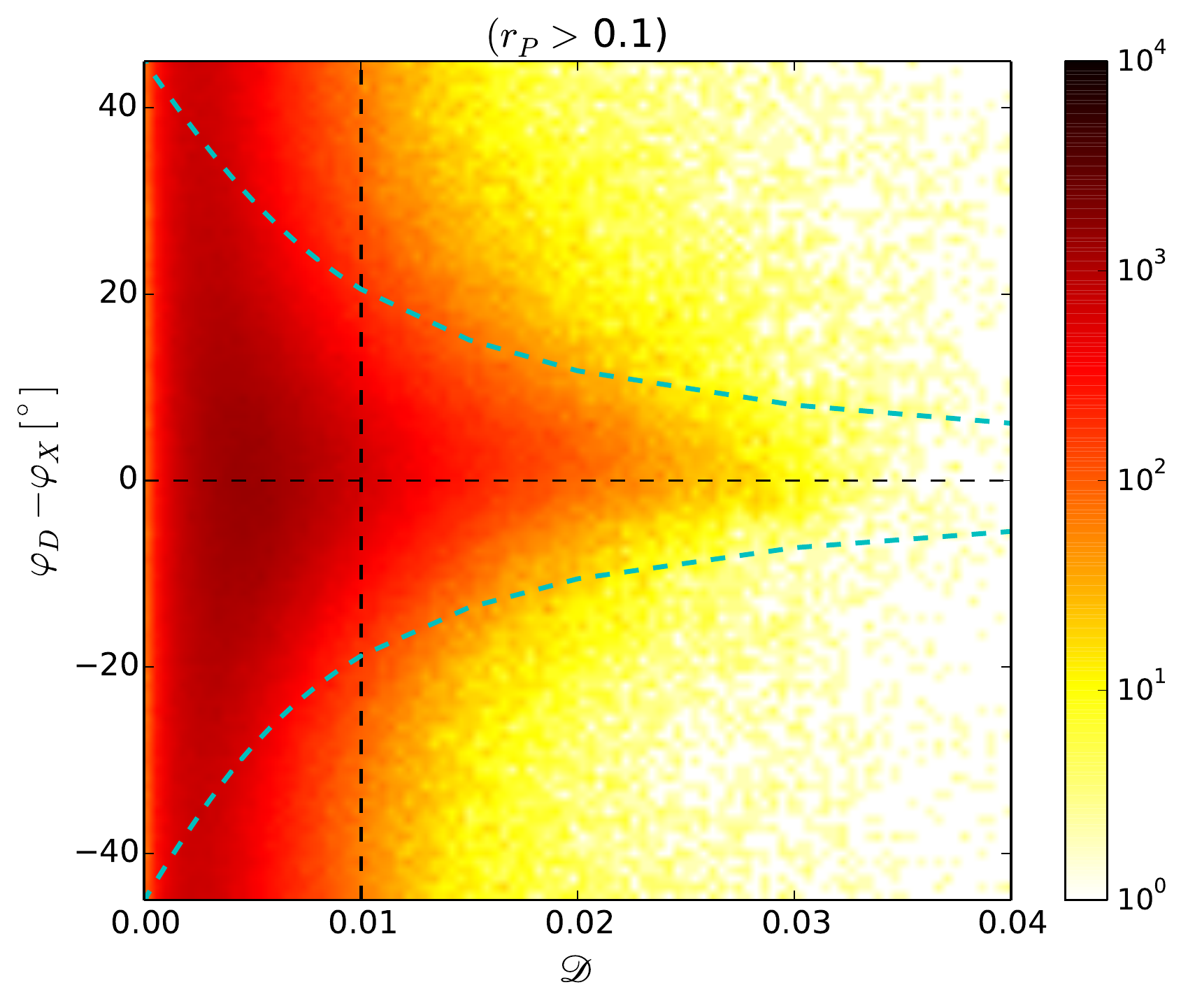} &
	\includegraphics[height=0.43\textwidth, trim={0 0 0 0}, clip]{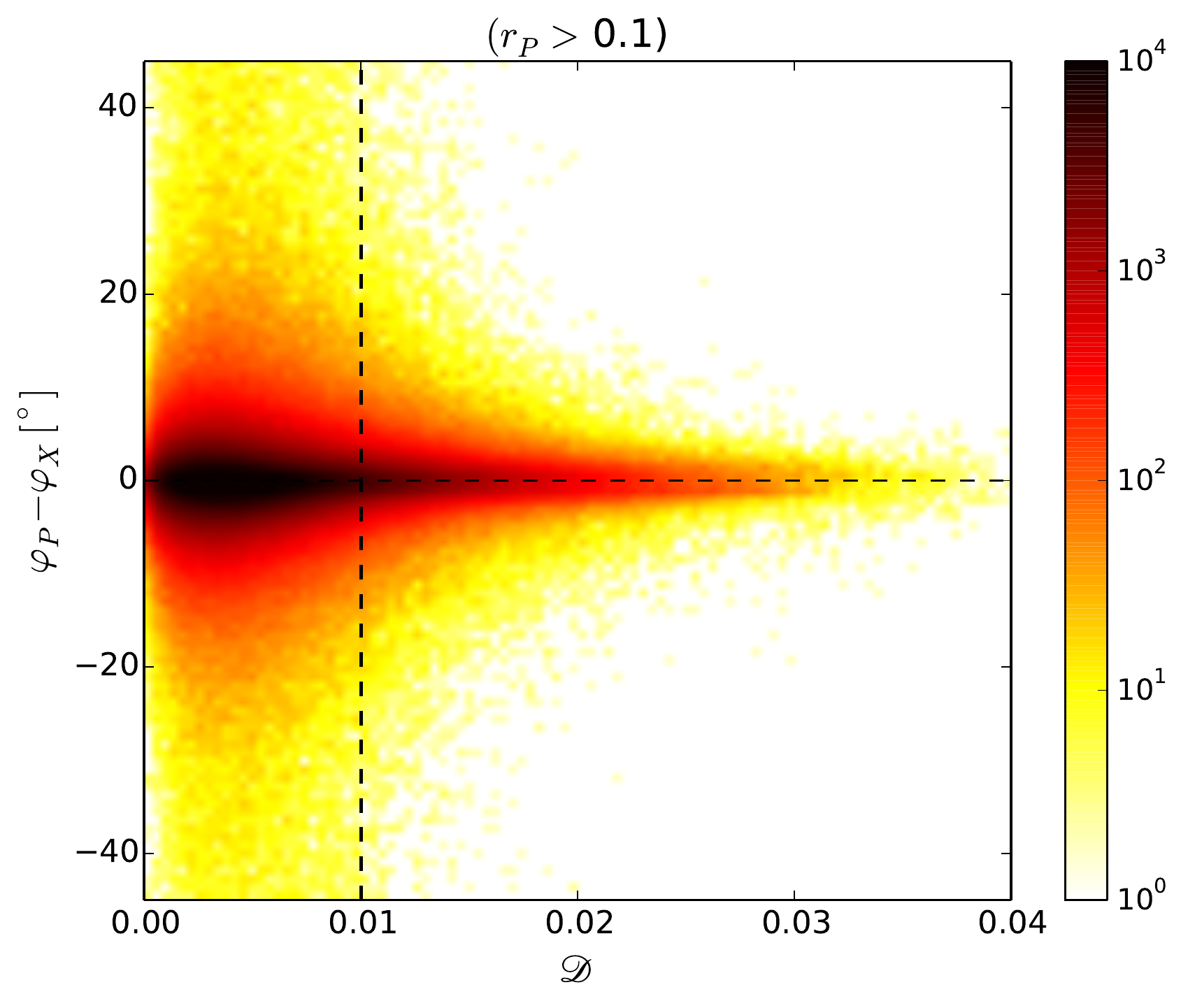} 
	\end{array}$
\caption{\textbf{(a)} 2D histogram showing the difference between the direction angle $\varphi_D$ derived from the DI measurement and the direction angle $\varphi_X$ derived from the XP measurement plotted against the measured diattenuation $\mathscr{D}$. The dashed cyan lines correspond to the maximum difference (pos./neg.) as predicted by the numerical study (see \cref{fig:Numerical-Diattenuation}d, for $\alpha=0^{\circ}$). \textbf{(b)} 2D histogram showing the difference between the direction angle $\varphi_P$ derived from the 3D-PLI measurement and $\varphi_X$ plotted against $\mathscr{D}$. To ensure a sufficient signal-to-noise ratio for $\varphi_X$ (cf.\ \cref{fig:DeltaDIR-XPvsRET}b), only regions with retardations $r_P > 0.1$ were selected for evaluation. The number of bins in the 2D histograms is 100 for both axes, respectively. The dashed vertical lines mark the region ($\mathscr{D}>1\,\%$) for which the diattenuation signal is expected to be mainly caused by the brain tissue and not by non-ideal system components (cf.\ \cref{sec:Diattenuation_Numerical}).}
\label{fig:DeltaDIR-XPvsDIA}
\end{figure}

To investigate how the difference between the measured direction angles ($\varphi_D$, $\varphi_P$) and the actual fiber direction angle (represented by $\varphi_X$) changes with the strength of the diattenuation signal, $(\varphi_D - \varphi_X)$ and $(\varphi_P - \varphi_X$) were plotted against $\mathscr{D}$ for regions with retardation values $r_P > 0.1$ (see \cref{fig:DeltaDIR-XPvsDIA}). In \cref{sec:DeltaDIR_vs_RET}, the same differences are plotted against $r_P$ for regions with diattenuation $\mathscr{D}>1\,\%$. 

As can be seen in \cref{fig:DeltaDIR-XPvsDIA}a, the direction angle $\varphi_D$ is broadly distributed around the actual fiber direction described by $\varphi_X$. The direction angle $\varphi_P$ is less broadly distributed (see \cref{fig:DeltaDIR-XPvsDIA}b). For $75\,\%$ of the values, ($\varphi_P - \varphi_X$) lies within $[-3.2^{\circ}, 2.49^{\circ}]$.
With increasing diattenuation, the mean absolute differences become smaller. 

\begin{figure}[htbp]
\centering
\includegraphics[width=0.6 \textwidth]{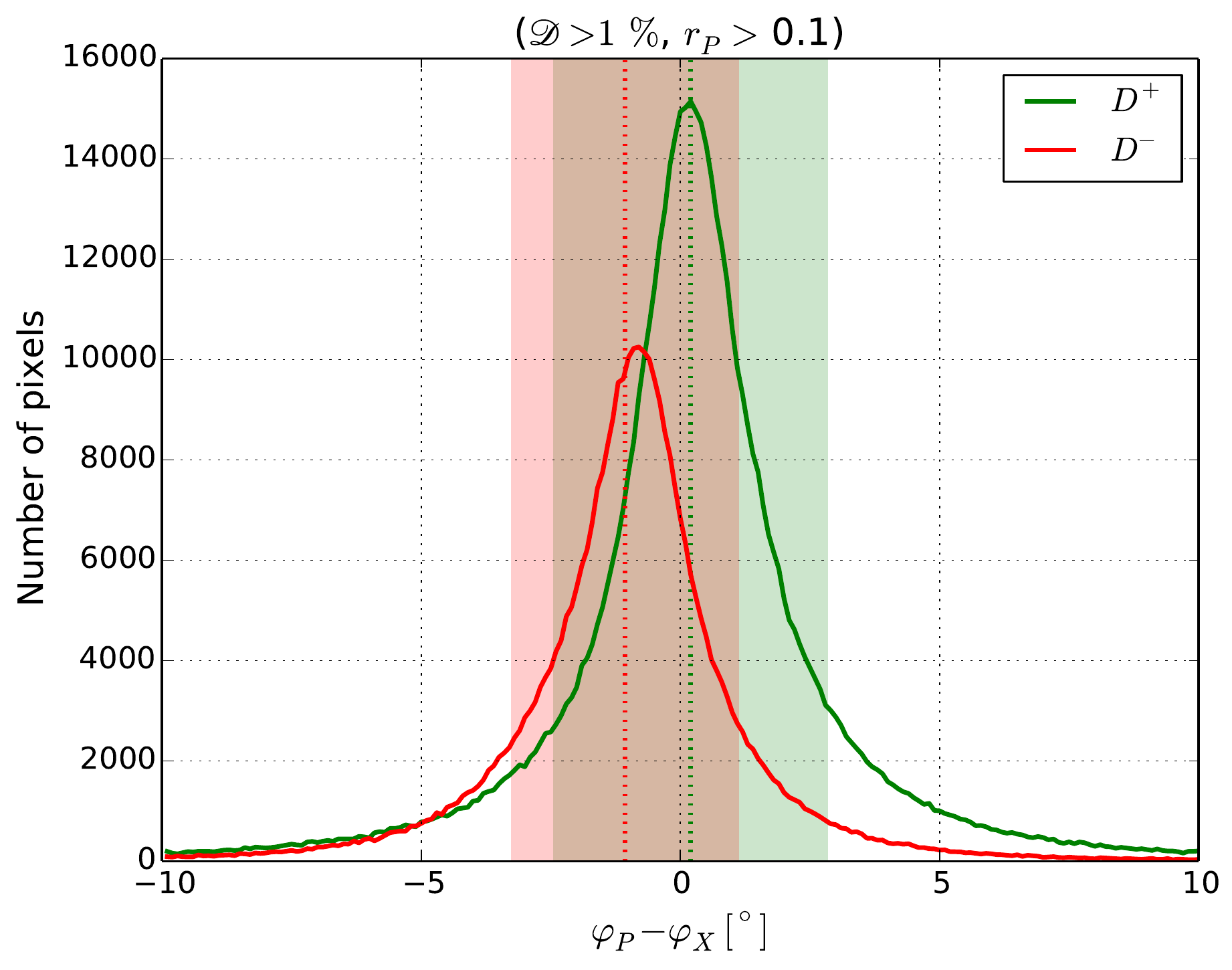}
\caption{Histogram showing the difference between the direction angle $\varphi_P$ derived from the 3D-PLI measurement and the direction angle $\varphi_X$ derived from the XP measurement computed for regions with diattenuation of type $D^+$ (green) and $D^-$ (red) according to \cref{fig:Maps_+-D} (bin width $= 0.1^{\circ}$). The green and red dashed lines indicate the respective mean values: $0.19^{\circ}$ for $D^+$ and $-1.07^{\circ}$ for $D^-$. The highlighted areas contain $75\,\%$ of the respective data: $[-2.45^{\circ}, 2.84^{\circ}]$ for $D^+$ and $[-3.27^{\circ}, 1.13^{\circ}]$ for $D^-$. To ensure a sufficient signal-to-noise ratio for $\varphi_X$ (cf.\ \cref{fig:DeltaDIR-XPvsRET}b), only regions with retardations $r_P > 0.1$ were selected for evaluation.}
\label{fig:Histo_DeltaPhi_PLI-XP2}
\end{figure}

To investigate the impact of diattenuation on the measured 3D-PLI signal in more detail, histograms of $(\varphi_P - \varphi_X)$ were computed separately for regions with diattenuation of types $D^+$ and $D^-$ (see \cref{fig:Histo_DeltaPhi_PLI-XP2}). The mean difference between the direction angles is $0.19^{\circ}$ for regions with $D^{+}$ and $-1.07^{\circ}$ for regions with $D^{-}$. Thus, on average, the fiber direction angle is slightly over-estimated in regions with $D^{+}$ and under-estimated in regions with $D^{-}$. For $75\,\%$ of the selected pixels, the difference lies within $[-2.45^{\circ},2.84^{\circ}]$ for regions with $D^+$ and within $[-3.27^{\circ},1.13^{\circ}]$ for regions with $D^-$.


\subsection{Discussion}
\label{sec:Discussion}

The experimental study quantified the diattenuation of brain tissue and its impact on 3D-PLI, and demonstrated that the diattenuation signal reveals additional structural information about the brain tissue.

Overall, the diattenuation of the investigated brain sections was found to be small ($\mathscr{D} < 4 \,\%$). The order of magnitude is the same as reported in other biological tissues \cite{ghosh2011,westphal2016,jiao2002,park2004,todorovic2004,martin2013}. 

The measured diattenuation and retardation signals showed no distinct correlation (see \cref{sec:Correlation_between_diattenuation_retardation}). A possible explanation could be that the diattenuation is not only caused by anisotropic absorption (dichroism) but also by scattering of light.
In regions where the diattenuation is only caused by dichroism, the measured diattenuation signal is expected to be proportional to the measured retardation signal, while the amount of (non-linear) scattering is not necessarily related to the strength of birefringence (see \cref{sec:Diattenuation}). Thus, scattering-induced diattenuation seems to play a major role and should be further investigated, e.\,g.\ by means of finite-difference time-domain simulations as described in \cite{menzel2016}. 

The experimental study has shown that the diattenuation has only a small impact on the fiber orientations determined in 3D-PLI. Figures \ref{fig:DeltaDIR-XPvsDIA}b and \ref{fig:Histo_DeltaPhi_PLI-XP2} show that the deviation of the determined direction angle $\varphi_P$ from the actual fiber direction (represented by $\varphi_X$) is less than $3^{\circ}$ for about $75\,\%$ of the analyzed pixels, independently of the type of diattenuation. Thus, the diattenuation of brain tissue can be neglected in the 3D-PLI analysis.

Due to the low signal-to-noise ratio of the diattenuation signal, the direction angle $\varphi_D$ determined from the DI measurement is more error-prone than the direction angle $\varphi_P$ determined from the 3D-PLI measurement (the distribution in \cref{fig:DeltaDIR-XPvsDIA}a is broader than in \cref{fig:DeltaDIR-XPvsDIA}b). With increasing tissue diattenuation, the mean absolute differences between the determined direction angles ($\varphi_D$ and $\varphi_P$) and the actual fiber direction (represented by $\varphi_X$) decrease. 

Despite the broad distribution of $\varphi_D$, the extremal transmittance is clearly correlated with the orientation of the nerve fibers. In some brain regions (e.\,g.\ within the ventral hippocampal comissure or the optic tract), the transmitted light intensity becomes maximal if the light is polarized \textit{parallel} to the fiber axes (described by $D^+$ in \cref{fig:Maps_+-D}). In other brain regions (e.\,g.\ within the caudate putamen), the transmitted light intensity becomes maximal if the light is polarized \textit{perpendicularly} to the fiber axes (described by $D^-$ in \cref{fig:Maps_+-D}). 
Presumably, differences in the fiber composition or structure of these brain regions exist that cause the polarized light to be attenuated differently when passing through the brain tissue.
These structural differences could concern the molecular composition of the fibers (more/less myelin, different lipid composition, etc.) as well as the macroscopic architecture of the fibers (long-range inter-cortical versus branching fibers, different inter-fiber distances, etc.). 
Dedicated structural tissue studies are required to confirm these hypotheses.


\section{Overall Discussion}
\label{sec:Overall-Discussion}

Comparing the findings of the numerical and experimental study (\cref{sec:Numerical_Study,sec:Experimental_Study_on_Brain_Tissue}) shows that the general predictions of the numerical study could be confirmed in the experimental study.

The difference between the direction angle $\varphi_D$ determined from the DI measurement and the actual fiber direction (represented by $\varphi_X$ determined from the XP measurement) behaves as predicted by the numerical study (\cref{sec:Diattenuation_Numerical}): The mean absolute difference is large for regions with diattenuations $\mathscr{D} \leq 1\,\%$ and decreases with increasing diattenuation. The measured values for $(\varphi_D - \varphi_X)$ lie mostly between the dashed cyan lines in \cref{fig:DeltaDIR-XPvsDIA}a, which indicate the maximum (positive/negative) difference between $\varphi_D$ and $\varphi$ as computed in the numerical study (cf.\ \cref{fig:Numerical-Diattenuation}d).

The numerical study (\cref{sec:3DPLI_Numerical}) also predicted a small impact of the tissue diattenuation on the direction angle $\varphi_P$ determined from the 3D-PLI measurement, which was confirmed in the experimental study (see Figs.\ \ref{fig:DeltaDIR-XPvsDIA}b and \ref{fig:Histo_DeltaPhi_PLI-XP2}).

The difference between the direction angles $\varphi_P$ and $\varphi_X$ shows a different pattern than for $\varphi_D$ (see \cref{fig:DeltaDIR-XPvsDIA}b). In regions with almost zero diattenuation, ($\varphi_P - \varphi_X$) is very small. With increasing diattenuation ($0 < \mathscr{D} < 0.5\,\%$), the distribution broadens rapidly. For $\mathscr{D}>0.5\,\%$, the distribution narrows with increasing diattenuation. This is in accordance with the predictions of the numerical study (\cref{sec:3DPLI_Numerical}d), taking into account that lower diattenuation values are expected for steep fibers (cf.\ \cref{sec:Numerical_Discussion}): In regions with zero diattenuation, $\varphi_P$ is equal to $\varphi$ for all inclinations. In regions with small diattenuation values, contributions from fibers of all inclinations are expected, i.\,e.\ a broad distribution of ($\varphi_P - \varphi$). With increasing diattenuation, less contributions from steep fibers are expected, which causes the difference between $\varphi_P$ and $\varphi$ to become smaller.

As mentioned in \cref{sec:Muller-Stokes_Calculus}, the diattenuation effect in regions with $D^-$ can be described by replacing the variable $D$ in all analytical formulas by the variable $(-D)$, which causes the simulated graphs in Figs.\ \ref{fig:Numerical_3D-PLI}d and \ref{fig:Numerical-Diattenuation}d to be mirrored along the x-axis. For \cref{fig:Numerical-Diattenuation}d, this makes almost no difference because the graphs are almost symmetric with respect to the x-axis. For \cref{fig:Numerical_3D-PLI}d, the simulated graphs $(\varphi_P - \varphi)$ are positive for $D^+$ and negative for $D^-$. This corresponds to the experimental observation that the histogram of $(\varphi_P - \varphi_X)$ shows a positive mean for regions with $D^+$ and a negative mean for regions with $D^-$ (see \cref{fig:Histo_DeltaPhi_PLI-XP2}).

Thus, the experimental results are in good accordance with the numerical results when taking into account two different types of diattenuation ($D^+$ and $D^-$). 
The study design combining the analytical description and system characterization with a numerical and experimental study enabled to measure and quantify the small diattenuation signal of brain tissue. The introduced procedures can be used as a routine for further measurements and are also applicable to other optical systems. The experimental study was performed exemplary on five sagittal rat brain sections and serves as a proof of principle that the developed model can be used to make general predictions. To further validate the theoretical predictions and to better understand how the DI measurement can help to reveal structural information, more extensive studies on various brain regions and species are needed.


\section{Conclusion}
\label{sec:Conclusion}

In this study, the diattenuation of brain tissue and its impact on 3D-Polarized Light Imaging were investigated for the first time. 
The diattenuation effect was explored analytically as well as in numerical and experimental studies by performing three different types of measurements: 3D-Polarized Light Imaging (3D-PLI), crossed polars (XP) measurement, and Diattenuation Imaging (DI).

A complete analytical description of the employed optical system was provided, including filter properties and tissue diattenuation using the Müller-Stokes calculus. Based on a thorough characterization of the employed polarimeter, the impact of the non-ideal system components on the tissue measurements was estimated in a numerical study. As the experimental results were in accordance with the findings of the numerical study, the analytical description and the determined system parameters can be used to model the experiment and to make general predictions. In addition, calibration procedures for the 3D-PLI, XP, and DI measurements were developed and their applicability was demonstrated. These characterization and calibration procedures can also be applied to other polarimetric systems.

The issues raised in the introduction can be answered as follows: (a) The diattenuation of the investigated brain sections is relatively small (less than $4\,\%$ for $60$\,\textmu m thin sections and the current preparation procedure). (b) The impact of diattenuation on the measured 3D-PLI signal was shown to be negligible. The small tissue diattenuation also implies that induced differences in the transmittance (e.\,g. between horizontal and vertical fibers) are expected to be small.
(c) Although the diattenuation effect has practically no impact on the measured 3D-PLI signal, the diattenuation measurement proved to be a valuable extension to 3D-PLI. By comparing the fiber directions extracted from 3D-PLI and DI, two different types of diattenuation effects can be distinguished that seem to be specific to certain anatomical regions and structures of the investigated brain sections. The phase and amplitude of the measured diattenuation signal can be used as imaging modalities providing different contrasts and structural information in addition to those obtained with 3D-PLI. Thus, Diattenuation Imaging is a promising imaging technique and reveals different types of fibrous structures that cannot be distinguished with current imaging techniques. The purpose of DI as additional imaging technique should be further investigated in the future.


\section*{Authors' Contributions}
M.M.\ contributed to the conception and design of the study. She provided the theoretical background, carried out the analytical calculations and the numerical study, computed the polarization-independent inhomogeneities and polarization properties of the optical system, performed the calibration of the XP measurement, participated in the evaluation and interpretation of the experimental data, and drafted the majority of the manuscript.
J.R.\ contributed to the conception and design of the study. She conducted the experimental measurements and the calibration of the 3D-PLI and DI measurements, performed the quality assurance of the experimental data, participated in the characterization of the optical system, carried out the analysis and evaluation of the experimental results, and drafted the experimental parts of the manuscript.
D.W.\ contributed to the characterization of the optical system, provided the optimization algorithm for fitting the polarization properties, and revised the manuscript.
H.K.\ developed the measurement protocol for the DI measurement and performed the tissue measurements and part of the filter measurements.
K.A.\ contributed to the anatomical content and the interpretation of results, and helped drafting the manuscript.
M.A.\ oversees the study, participated in the conception and design, contributed to the interpretation of the numerical and experimental results, and helped drafting the manuscript.
All authors read the final manuscript and gave final approval for publication.


\section*{Conflicts of Interests}
The authors declare that they have no relevant financial interests in the manuscript and no other potential conflicts of interest to disclose.


\section*{Funding}
This project received funding from the Helmholtz Association port-folio theme ‘Supercomputing and Modeling for the Human Brain’ and the European Union’s Horizon 2020 Research and Innovation Programme under Grant Agreement No.\ 7202070 (Human Brain Project SGA1). 


\section*{Acknowledgements}
We would like to thank M.\ Cremer, Ch.\ Schramm, and P.\ Nysten for the preparation of the histological rat brain sections.
Furthermore, we thank H. Wiese for helpful discussions about the calibration and characterization of the optical system, N. Schubert for the image registration, M. Schober for assistance in determining and correcting the parallax effects, and S. Köhnen for assistance in determining the rotation center of the images. Finally, we would like to thank K. Michielsen, H. de Raedt, and F. Matuschke for helpful discussions and advice.



\appendix


\section*{Appendix}

\section{Characterization of the Polarimeter}
\label{sec:Characterization_LAP}

It has been observed that calibration measurements (without brain tissue) yield a diattenuation signal that is of a similar order of magnitude as the measured diattenuation signal of brain tissue. To separate the effects induced by non-ideal optical components of the polarimeter from the tissue properties to be investigated, all components of the polarimeter that might have an impact on the measured diattenuation signal were characterized in a separate study.

The non-ideal optical properties of the polarimeter can be classified into polarization-independent inhomogeneities and polarization properties.

One source for polarization-independent inhomogeneities is a non-uniform illumination of the light source. Filter inhomogeneities introduced in the fabrication process can also cause inhomogeneities due to a non-uniform absorption of the light. In addition, the sensitivity of the CCD camera chip might not be exactly the same for all detector pixels. In this study (\cref{sec:Polarization-Independent_Inhomogeneities,sec:Characterization_Pol-Ind-Inhom}), the polarization-independent inhomogeneities were investigated in separate measurements of the light source and the filters (polarizer, retarder, analyzer). 

The polarization properties of the optical components are also expected to be non-ideal.
The light emitted by the LED light source might not be completely unpolarized and the sensitivity of the camera might depend on the polarization of the incoming light. Furthermore, the degree of polarization of polarizer and analyzer is expected to be less then $100\,\%$.
Previous studies also showed that the working wavelength of the employed quarter-wave retarder is not optimally adapted to the illumination wavelength of the LED light source which induces a phase retardation that is not exactly equal to $\pi/2$ \cite{reckfort2015}.
In this study (\cref{sec:Polarization_Properties,sec:Characterization_Pol-Prop}), the partially polarization of the light source, the polarization sensitivity of the camera, and the non-ideal polarization properties of the filters were estimated by fitting the polarization properties to the intensity curves obtained in various filter measurements.


\subsection{Methods}

\subsubsection{Optical Components of the Polarimeter}
\label{sec:Components_LAP}

The employed polarimeter\footnote{In previous publications, the polarimeter was referred to as \textit{Large-Area Polarimeter (LAP)} \cite{MAxer2011_1,MAxer2011_2,menzel2015,reckfort2015}.} consists (from bottom to top) of a light source, a linear polarizer, a quarter-wave retarder, a specimen stage, and a second linear polarizer (analyzer), see \cref{fig:Setups}a. Each filter can be rotated individually or be removed from the imaging system.

The customized light source (\textit{FZJ SSQ300-ALK-G} provided by \textit{iiM}, Germany) contains a matrix of LED diodes which illuminates an area of approximately $300 \times 300$\,mm$^2$. The emitted light is expected to be incoherent and unpolarized with a wavelength of $(525 \pm 25)\,$nm \cite{reckfort2015}. To create a more uniform illumination, diffuser plates are placed on top of the LED panel. 

The employed camera (\textit{AxioCam HRc} by \textit{Zeiss}) consists of a CCD sensor which uses a microscanning procedure to achieve higher spatial resolutions. The used image matrix is $4164 \times 3120$\,px$^2$. The camera is equipped with a lens (\textit{Rodenstock Apo-Rodagon-N}, 1:4, $f=90$\,mm, combined with \textit{Linos Modular Focus}) to achieve a higher optical resolution for the investigation of rodent brains (as compared to previous studies of primate brains \cite{MAxer2011_1,MAxer2011_2,reckfort2015}). The resulting object space resolution is 14\,\textmu m/px  with a field of view of approximately $58 \times 44$\,mm$^2$.

The employed linear polarizers (\textit{XP38}) and the quarter-wave retarder (\textit{WP140}) were manufactured by \textit{ITOS}, Germany. They are $239\,$mm in diameter and consist of polymer foils. 


\subsubsection{Polarization-Independent Inhomogeneities}
\label{sec:Polarization-Independent_Inhomogeneities}

To study the polarization-independent inhomogeneities, images of the light source and each filter were acquired in separate measurements. For all measurements, the same exposure time was used.

As the effects of light source and camera cannot be separated, they were investigated conjointly. To determine the inhomogeneities caused by an inhomogeneous illumination and a varying sensitivity of the camera pixels, all filters were removed from the setup and an image of the light source was acquired. The photon noise was reduced by repeating the measurement $50$ times and averaging the intensity for each image pixel. 

The filter inhomogeneities (i.\,e. the inhomogeneous light absorption of the polymer foils) were measured separately for each filter by removing the other filters from the imaging system and rotating the filter by $18$ discrete rotation angles in counter-clockwise direction. To reduce photon noise, each measurement was repeated $20$ times. To separate the inhomogeneities caused by light source and camera from the inhomogeneities of the filters, every acquired filter image was divided by the averaged image of the light source. Any remaining pixelated structure was compensated by applying a Gaussian filter with an adequate kernel size. To avoid an influence of a temporal light intensity change of the different LED diodes, the measurement of the light source (as described above) was performed just before every filter measurement.

To separate the light attenuation induced by the filter inhomogeneities from the light attenuation induced by polarization effects (e.\,g.\ a partially polarized light source), the images of each image series ($\rho = \{0^{\circ}, 10^{\circ}, \ldots, 170^{\circ} \}$) were digitally rotated in clockwise direction by the corresponding rotation angle $\rho$ so that all images have the same virtual angle position ($\rho=0^{\circ}$). For this purpose, the rotation center of the images was determined by identifying in each image series the point that is rotationally invariant. Then, the images were cropped to a circular region around the determined rotation center with a maximum possible diameter of $2914$\,px and the images were rotated by applying a bilinear interpolation. The resulting $18$ images were averaged pixel-wise for each image series.

As a measure for the inhomogeneities of light source and filters, the maximum and minimum intensity values of the averaged images were determined and the image contrast was computed via: $\mathcal{C} = (I_{\text{max}} - I_{\text{min}})/(I_{\text{max}} + I_{\text{min}})$, respectively.
The average transmittance of the filters was computed via: $\tau = (I_{\text{max}} + I_{\text{min}})/2$.
Note that $I_{\text{max}}$ and $I_{\text{min}}$ are here the maximum and minimum intensity values of the image and should not be confused with the maximum and minimum transmitted light intensities observed in a DI measurement.


\subsubsection{Polarization Properties}
\label{sec:Polarization_Properties}

In contrast to the polarization-independent inhomogeneities, a pixel-wise determination of the polarization parameters is not possible because the polarization effects are non-multiplicative and influence each other. For this reason, the polarization properties of the optical elements could only be determined as an average over the field of view. In principle, the polarization properties could be determined pixel-wise by using a high-quality reference (light source or filters). However, high-quality filters of comparable sizes are commercially not available and illuminating such large-area filters with a highly polarized light source (e.\,g.\ laser) is not possible with reasonable experimental effort.

The study was executed in two steps: First, the polarization properties were measured with different filter setups. Second, the polarization parameters were estimated by fitting intensity profiles obtained by modeling light source, camera, and filters as generalized Stokes vectors and Müller matrices to the measured intensity profiles.

\vspace{2mm} \noindent
{\textit{1) Filter Measurements:}} \,
The filter measurements were performed using different combinations of polarizer, retarder, and analyzer. Some of the filters were rotated by discrete rotation angles $\rho = \{0^{\circ}, 10^{\circ}, \ldots, 170^{\circ} \}$, while other filters were kept at a fixed angle position. Table \ref{tab:FilterMeasurements} lists all executed measurements and the respective angle positions of the filters.
\begin{table}
\begin{center}
  \begin{tabular}{l||c|c|c }
      			& Polarizer: $P_x(\xi = \rho)$	&	Retarder: $\Lambda(\xi = \rho - 45^{\circ})$ 	&	Analyzer: $P_y(\xi = \rho+90^{\circ})$ \\ \hline \hline
	$P_x$						& $\rho = \{0^{\circ}, 10^{\circ}, \ldots, 170^{\circ} \}$ & -- & -- \\ \hline
	$P_x \,P_y$					& $\rho = \{0^{\circ}, 10^{\circ}, \ldots, 170^{\circ} \}$ & -- & $\rho = \{270^{\circ}, 280^{\circ}, \ldots, 90^{\circ} \}$ \\ \hline
	$P_x \,\Lambda$				& $\rho = \{0^{\circ}, 10^{\circ}, \ldots, 170^{\circ} \}$ & $\rho = \{0^{\circ}, 10^{\circ}, \ldots, 170^{\circ} \}$ & -- \\ \hline
	$\Lambda \,P_y$				& -- & $\rho = \{0^{\circ}, 10^{\circ}, \ldots, 170^{\circ} \}$ & $\rho = \{270^{\circ}, 280^{\circ}, \ldots, 90^{\circ} \}$ \\ \hline
	$\Lambda \,P_y(270^{\circ})$	& -- & $\rho = \{0^{\circ}, 10^{\circ}, \ldots, 170^{\circ} \}$ & $\rho = 270^{\circ}$ \\ \hline \hline
	$P_x \,\Lambda(0^{\circ})$	& $\rho = \{0^{\circ}, 10^{\circ}, \ldots, 170^{\circ} \}$ & $\rho = 0^{\circ}$ & -- \\ \hline
	$P_x \,\Lambda(10^{\circ})$	& $\rho = \{0^{\circ}, 10^{\circ}, \ldots, 170^{\circ} \}$ & $\rho = 10^{\circ}$ & -- \\ \hline
	$P_x \,\Lambda(30^{\circ})$	& $\rho = \{0^{\circ}, 10^{\circ}, \ldots, 170^{\circ} \}$ & $\rho = 30^{\circ}$ & -- \\ \hline
	$P_x \,\Lambda(60^{\circ})$	& $\rho = \{0^{\circ}, 10^{\circ}, \ldots, 170^{\circ} \}$ & $\rho = 60^{\circ}$ & -- \\ \hline
	$P_x \,\Lambda(80^{\circ})$	& $\rho = \{0^{\circ}, 10^{\circ}, \ldots, 170^{\circ} \}$ & $\rho = 80^{\circ}$ & -- \\ \hline \hline	
	$\Lambda(0^{\circ})  \,P_y$	& -- & $\rho = 0^{\circ}$ & $\rho = \{270^{\circ}, 280^{\circ}, \ldots, 90^{\circ} \}$\\ \hline
	$\Lambda(10^{\circ}) \,P_y$	& -- & $\rho = 10^{\circ}$ & $\rho = \{270^{\circ}, 280^{\circ}, \ldots, 90^{\circ} \}$\\ \hline
	$\Lambda(30^{\circ}) \,P_y$	& -- & $\rho = 30^{\circ}$ & $\rho = \{270^{\circ}, 280^{\circ}, \ldots, 90^{\circ} \}$ \\ \hline
	$\Lambda(60^{\circ}) \,P_y$	& -- & $\rho = 60^{\circ}$ & $\rho = \{270^{\circ}, 280^{\circ}, \ldots, 90^{\circ} \}$ \\ \hline
	$\Lambda(80^{\circ}) \,P_y$	& -- & $\rho = 80^{\circ}$ & $\rho = \{270^{\circ}, 280^{\circ}, \ldots, 90^{\circ} \}$ \\ \hline \hline
	$P_x(0^{\circ}) \, P_y$	 	& $\rho = 0^{\circ}$ & -- & $\rho = \{270^{\circ}, 280^{\circ}, \ldots, 90^{\circ} \}$ \\ \hline
	$P_x \, P_y(270^{\circ})$		& $\rho = \{0^{\circ}, 10^{\circ}, \ldots, 170^{\circ} \}$ & -- & $\rho = 270^{\circ}$ \\ \hline 
	$P_x(0^{\circ}) \, \Lambda \, P_y(270^{\circ})$ & $\rho = 0^{\circ}$ & $\rho = \{0^{\circ}, 10^{\circ}, \ldots, 170^{\circ} \}$ & $\rho = 270^{\circ}$ 
	\\ \hline
	\end{tabular} 
  \caption{Configuration of the filter measurements to determine the polarization properties of the optical components: The angle position $\xi$ of the filters (polarizer/retarder/analyzer) is defined in terms of the rotation angle $\rho$ as described in \cref{sec:Measurement_Setups_Analysis} }
  \label{tab:FilterMeasurements} 
\end{center}
\end{table}
\enlargethispage{0.5cm}
All filter measurements were repeated three times, averaged, and divided by the (averaged) image of the light source, which was recorded $50$ times before every filter measurement.
To avoid fringe effects from absorbing elements at the image border, the intensity was averaged over a circular region with a diameter of $2914$\,px around the rotation center of the filters for each rotation angle $\rho$, and the standard deviation $\sigma(\rho)$ was determined. The resulting intensity values $I(\rho)$ were divided by the average intensity $\overline{I}$ over all rotation angles. As the intensities were averaged over the rotation center, the polarization-independent inhomogeneities (described in Sec.\ \ref{sec:Polarization-Independent_Inhomogeneities}) add the same attenuation to the signal for each rotation angle and could therefore be neglected when computing the normalized light intensity profiles.


\vspace{2mm} \noindent
{\textit{2) Fitting of Polarization Parameters:}} \,
As the polarization properties of light source and camera are unknown, they were modeled as generalized, normalized Stokes vectors ($\vec{S}_L$ and $\vec{S}_c$) as defined in \cref{eq:StokesVector2} with $I = 1$:
\begin{align}
\vec{S}_L
=
\begin{pmatrix} 1 \\
				p_L \,\, \cos({2 \, \psi_L})\,\cos({2 \, \chi_L}) \\
				p_L \,\, \sin({2 \, \psi_L})\,\cos({2 \, \chi_L}) \\
				p_L \,\, \sin({2 \, \chi_L})
\end{pmatrix}, \,\,\,\,\,\,\,
\vec{S}_c
=
\begin{pmatrix} 1 \\
				p_c \,\, \cos({2\,\psi_c})\,\cos({2\,\chi_c}) \\
				p_c \,\, \sin({2\,\psi_c})\,\cos({2\,\chi_c}) \\
				p_c \,\, \sin({2\,\chi_c})
\end{pmatrix}.
\label{eq:SL_Sc_definition}
\end{align}

The filters (polarizer, retarder, analyzer) were described by the Müller matrices ($P_x$, $\Lambda$, $P_y$) as defined in \cref{eq:Analyzer_non-ideal,eq:Retarder_non-ideal,eq:Polarizer_non-ideal}. To enable a direct comparison with the measured normalized light intensity profiles, the average transmittance of each optical element was set to one ($\tau_x = \tau_y = \tau_{\Lambda} = 1$). 

By multiplying the Stokes vectors with the corresponding Müller matrices (using the rotation angles specified in \cref{tab:FilterMeasurements}) and evaluating the first entry of the resulting Stokes vectors, the transmitted light intensity was computed for each filter measurement. The computations were performed using distributed computing on desktop computers consuming about $500$ core hours.

The modeled intensity profiles $I_{\text{model}}(\rho)$ were fitted to the measured intensity profiles $I_{\text{meas}}(\rho)$ for each rotation angle ($\rho = \{0^{\circ}, 10^{\circ}, \ldots, 170^{\circ} \}$) by minimizing the sum of squared differences $\Delta$ between the measured and the modeled intensity profiles. To account for random errors (e.\,g.\ read-out noise or a varying sensitivity between camera pixels), the differences were normalized by the determined standard deviation $\sigma(\rho)$. To ensure a fair weighting of high and low signal amplitudes, the differences were divided by the measured signal amplitude $(I_{\text{meas,max}} - I_{\text{meas,min}})$ for each of the $k$ filter measurements defined in \cref{tab:FilterMeasurements}:
\begin{align}
\Delta = \sum_{j,k} \left[ \frac{ \big(I_{\text{meas}}(\rho) - I_{\text{model}}(\rho) \big)^2}{ \sigma^2(\rho)\,\left(I_{\text{meas,max}} - I_{\text{meas,min}}\right)^2} \right]_k .
\end{align}

To find the global minimum of $\Delta$, the sum of squared differences was minimized numerically using a differential evolution algorithm \cite{storn1997}. 
The Stokes parameters of light source $\{ p_L, \psi_L, \chi_L\}$ and camera $\{p_c, \psi_c, \chi_c \}$ were fitted for $D_x, D_y = \{ 0.9,\, 0.905,\,0.91,\,\ldots,\,1 \}$ and $\gamma = \{ 0.4\,\pi, 0.405\,\pi, 0.41\,\pi, \ldots, 0.6\,\pi \} $. The parameters $D_x$, $D_y$, and $\gamma$ were not fitted to reduce computing time and to ensure a proper convergence of the algorithm. The parameter ranges were chosen such that the whole range of possible values was inclosed.


\subsection{Results and Discussion}

\subsubsection{Polarization-Independent Inhomogeneities}
\label{sec:Characterization_Pol-Ind-Inhom}

\begin{figure}[htbp]
\centering
\includegraphics[width=1 \textwidth]{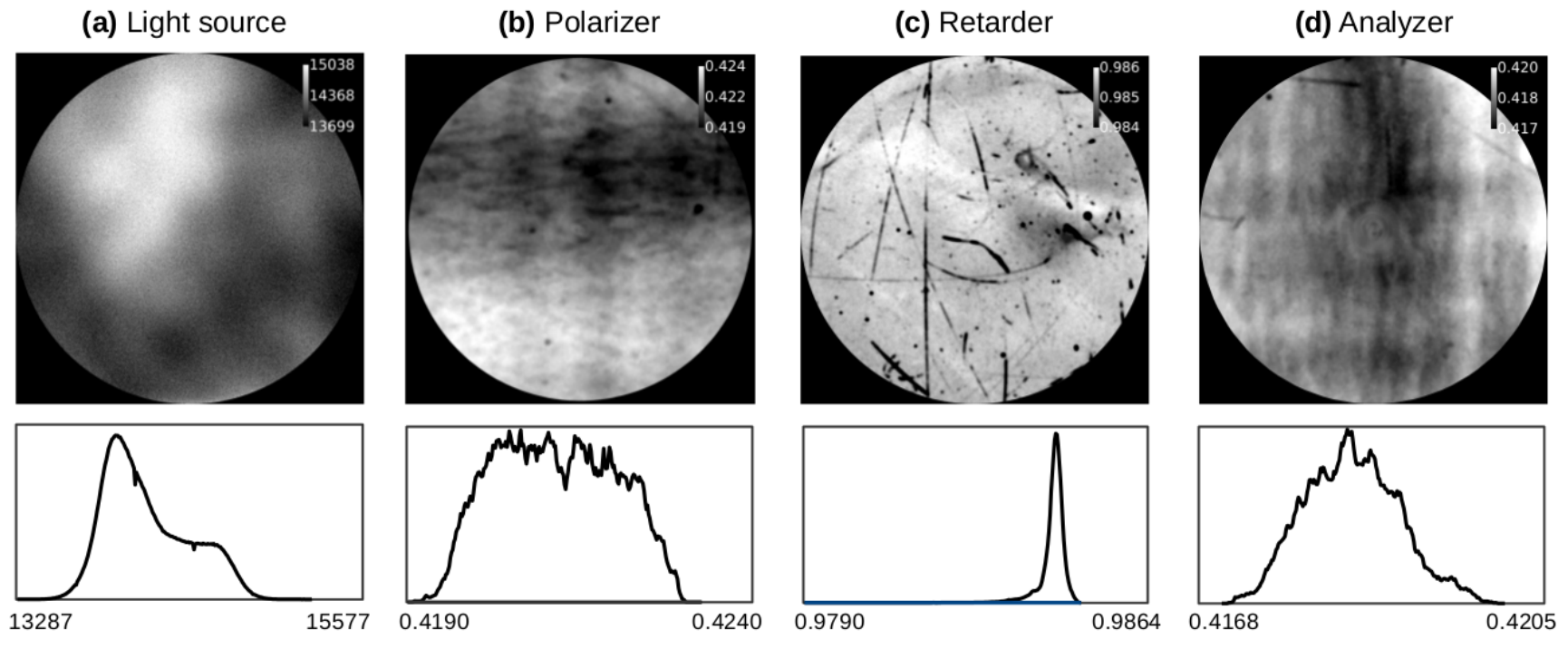}
\caption{Images and histograms of light source, polarizer, retarder, and analyzer. Note that the contrast of the images is different (the maximum measurable intensity values are depicted in white and the minimum values in black, respectively).}
\label{fig:Pol-Ind-Inhomogeneities}
\end{figure}

It has been observed that a pixelated structure remains in the background when dividing the filter images by the image of the light source. To compensate this effect, a Gaussian blur was applied to the normalized images with $\sigma = 10\,$px (for polarizer and analyzer) and $\sigma = 5\,$px (for the retarder). The resulting images and histograms of light source, polarizer, retarder, and analyzer are shown in \cref{fig:Pol-Ind-Inhomogeneities}. Dust particles have been excluded from the analysis.

The light source (see \cref{fig:Pol-Ind-Inhomogeneities}a) shows regions with varying light intensity. The maximum measurable contrast is about $8\,\%$. The polarizer and analyzer (see \cref{fig:Pol-Ind-Inhomogeneities}b,d) show patterns of large horizontal and vertical stripes, which are most likely caused by the stretching of the polymer foils during the fabrication process. The maximum measurable contrast is about $0.59\,\%$ for the polarizer and about $0.44\,\%$ for the analyzer. The average transmittances of polarizer and analyzer are about $42\,\%$ ($\tau_x \approx 0.422$, $\tau_y \approx 0.419$). The quarter-wave retarder (see \cref{fig:Pol-Ind-Inhomogeneities}c) shows several scratches, but is still quite homogeneous. The maximum measurable contrast is about $0.38\,\%$ and the average transmittance about $98$\,\% ($\tau_{\Lambda} \approx 0.983$).


\subsubsection{Polarization Properties}
\label{sec:Characterization_Pol-Prop}

\begin{figure}[htbp]
\centering
\includegraphics[width=1 \textwidth]{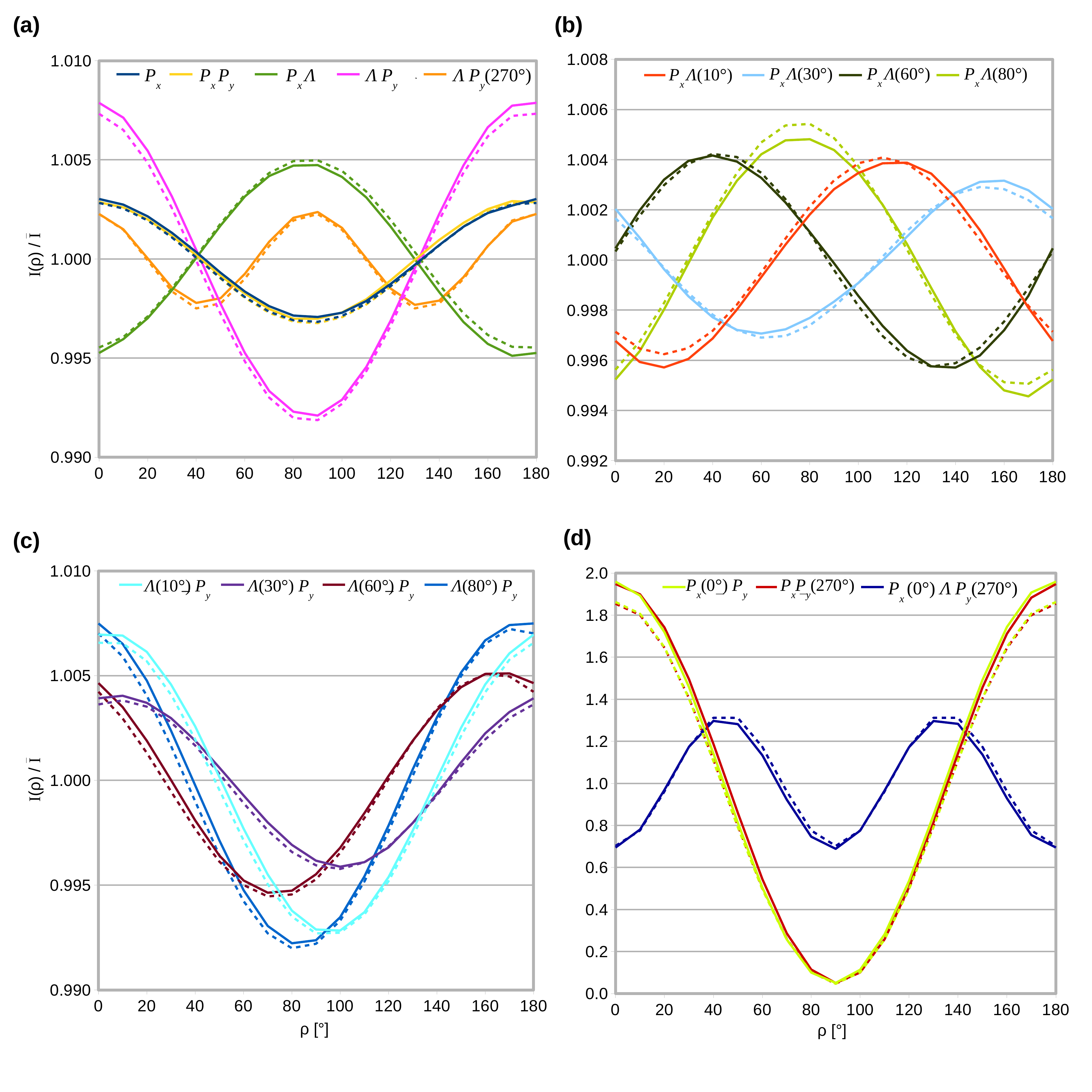}
\caption{Normalized light intensity profiles: The solid curves show the profiles obtained from the filter measurements described in \cref{tab:FilterMeasurements}. The dashed curves show the modeled profiles computed from the polarization parameters defined in \cref{tab:Polarization_Properties}, first row.}
\label{fig:ModelFit}
\end{figure}

Figure \ref{fig:ModelFit} shows the (normalized) light intensity profiles of the filter measurements (solid curves) as well as the modeled light intensity profiles (dashed curves) that were computed for each filter measurement from the determined polarization parameters for which the sum of squared differences $\Delta$ is minimized (see best fit in \cref{tab:Polarization_Properties}). The relative difference between the curves is mostly less than $5\,\%$ of the measured signal amplitude, which demonstrates that the fit of the polarization parameters is good.

Table \ref{tab:Polarization_Properties} shows the computed polarization parameters for the best fit $(\Delta_{\text{min}} = 2138.52)$ as well as the average, the standard deviation, and the relative standard deviation (divided by the average) of the polarization parameters belonging to the best 20 fits (with $\Delta = 2138.52, \ldots, 2154.68$). The relative standard deviation is mostly less than $1\,\%$, except for $\chi_L$ which is consistent with zero. This demonstrates that the fitted polarization parameters are stable and can be used as estimates to describe the polarization properties of light source, filters, and camera.
\begin{table}[htbp]
\begin{center}
  \begin{tabular}{|l||c|c|c|c|c|r|r|r|r|}
  \hline
& $D_x$ & $D_y$ & $\gamma$ / $\pi$ & $p_L$ & $\psi_L$ & $\chi_L$ \,\,\,\,\,\, & $p_c$ \,\,\,\,\,\, & $\psi_c$ \,\,\,\,\,\, & $\chi_c$ \,\,\,\,\,\, \\ \hline \hline
Best Fit & 0.98000 & 0.97000 & 0.49000 & 0.00513 & 1.49472 & $-5 \times 10^{-5}$ & 0.00822 & 3.07458 & -0.03068\\ \hline \hline
Average & 0.98075 & 0.96900 & 0.49000 & 0.00513 & 1.49469 & $3 \times 10^{-6}$ & 0.00822 & 3.07457 & -0.03065\\ \hline 
Std. Dev. & 0.00438 & 0.00369 & 0.00284 & 0.00006 & 0.00011 & $8 \times 10^{-5}$ & 0.00006 & 0.00004 & 0.00032\\ \hline 
Rel. Std. Dev. & 0.00446 & 0.00381 & 0.00579 & 0.01234 & 0.00007 & 32\quad\quad & 0.00738 & 0.00001 & 0.01053\\ \hline 
	\end{tabular} 
  \caption{Fitted polarization properties of filters, light source, and camera: The parameters in the first row minimize the sum of squared differences ($\Delta_{\text{min}} = 2138.52$). The other parameters show the average, standard deviation, and relative standard deviation (divided by the average) for the best $20$ fits with $\Delta = 2138.52, \ldots, 2154.68$.}
  \label{tab:Polarization_Properties} 
\end{center}
\end{table}

The sum of squared differences is minimized for $D_x = 0.98$, $D_y = 0.97$, and $\gamma = 0.49\,\pi$. Thus, the employed linear polarizers have a similar degree of polarization which is slightly less than $100\,\%$. The retardance of the quarter-wave retarder is slightly less than $\pi/2$ (probably due to the wavelength discrepancy between the light source and the retarder \cite{reckfort2015}). 

Inserting the best fit parameters $\{p_L, \psi_L, \chi_L \}$ and $\{p_c, \psi_c, \chi_c \}$ from \cref{tab:Polarization_Properties} into \cref{eq:SL_Sc_definition}, yields the following Stokes vectors for light source and camera:
\begin{align}
\vec{S}_L
\approx
\begin{pmatrix} 1 \\
				-5 \times 10^{-3} \\
				\,\,\,\,\,\, 8 \times 10^{-4} \\
				-5 \times 10^{-7}
\end{pmatrix}, \,\,\,\,\,\,\,\,\,\,\,
\vec{S}_c
\approx
\begin{pmatrix} 1 \\
				\,\,\,\,\,\, 8 \times 10^{-3} \\
				-1 \times 10^{-3} \\
				-5 \times 10^{-4}
\end{pmatrix}.
\label{eq:SL_Sc_parameters}
\end{align}
Hence, about $0.5\,\%$ of the ingoing light is vertically linearly polarized $\big(\arctan(S_{L,2}/S_{L,1})/2 \approx 86^{\circ}\big)$; the bigger part of the light is unpolarized. The camera has a small preference for horizontally linearly and left-handed circularly polarized light ($0.8\,\%$ linear polarization oriented at $6^{\circ}$ and $0.05\,\%$ left-handed circular polarization).


\subsection{Conclusion}
\label{sec:Characterization_Conclusion}

The inhomogeneous illumination of the light source ($8\,\%$ image contrast) is much more dominant than the polarization-independent inhomogeneities of the filters ($0.4 - 0.6\,\%$ image contrast) and needs to be taken into account when analyzing the tissue measurements. The inhomogeneities of light source and camera can be compensated by dividing the acquired tissue images by an image without specimen. To further compensate the inhomogeneities of the filters and the polarization effects of the optical components, a calibration as proposed by Dammers et al.\,\cite{dammers2011} can be performed.

For the XP measurement, an equivalent calibration is not possible because a measurement without specimen would result in almost zero light intensity due to the $90^{\circ}$-orientation of polarizer and analyzer. Instead, the images obtained from the XP measurement can be divided by the image of the light source and by the polarization-independent filter inhomogeneities of polarizer and analyzer (see \cref{fig:Pol-Ind-Inhomogeneities}a,b,d). As the filter inhomogeneities can only be determined for a circular region around the rotation center of the filters, the images obtained in the XP measurement need to be cropped to this region. A possible, significant parallax effect induced by the analyzer should be corrected before the calibration.

It should be noted that these calibration methods only correct for multiplicative errors in the measured light intensity (i.\,e.\ inhomogeneous illumination, absorption, detection sensitivity, etc.). Differences between presumed and actual polarization states (e.\,g.\ induced by a partially polarized light source, non-ideal polarization filters, or a polarization-sensitive camera as described in \cref{sec:Characterization_Pol-Prop}) are not taken into account. When analyzing the data, these effects need to be considered as described in \cref{sec:Numerical_Study} and \cref{fig:Numerical-Diattenuation,fig:Numerical_3D-PLI}.


\section{Correction of the 3D-PLI Fiber Orientations}
\label{sec:Corr_Fiber_Orientation}

In principle, the impact of the tissue diattenuation and the non-ideal filter properties on the fiber orientations ($\varphi_P$, $\alpha_P$) derived from 3D-PLI can be corrected exactly.
If the values for $D_x$, $D_y$, $\gamma$, and $D$ were known pixel-wise, the actual retardance $\delta$ of the brain tissue could be computed from the Fourier coefficients in \cref{eq:a0,eq:a2,eq:b2}:
\begin{align}
\delta &= \arccos \left( \frac{E \pm \sqrt{E^2 + 4(F-G)(H-G)}} {2(G-H)} \right), 
\label{eq:delta_corr}
\end{align}
\begin{align}
&E \equiv \frac{a_{2P}^2 + b_{2P}^2}{a_{0P}^2} \, \cos\gamma \,\,\, D_x \, D_y \, \sqrt{1-D^2} \, \left( 1 - \frac{1}{2} \cos\gamma \,\,\, D_x \, D_y \right), 
&&G \equiv \sin^2\gamma \,\,\, D_x^2 \,\, D_y^2 \,\, \big(1 - D^2 \big), \\
&F \equiv \frac{a_{2P}^2 + b_{2P}^2}{a_{0P}^2} \left( 1 - \frac{1}{2} \cos\gamma \,\,\, D_x \, D_y \right)^2 - \,D^2 \big( \cos\gamma \,\,\, D_x - D_y \big)^2, 
&&H \equiv - \frac{a_{2P}^2 + b_{2P}^2}{4 \, a_{0P}^2} \, \cos^2\gamma \,\,\, D_x^2 \,\, D_y^2 \,\, \big(1 - D^2 \big).
\end{align}
Using the determined value for $\delta$, the actual fiber inclination angle $\alpha$ could be computed via \cref{eq:delta_approx,eq:alpha_corr}. 
The actual fiber direction angle $\varphi$ could be computed using $\delta$ and \cref{eq:a2,eq:b2}. However, a pixel-wise measurement of the filter properties is not feasible (see \cref{sec:Polarization_Properties}) and therefore this correction is not used in the 3D-PLI analysis.


\section{Parameter Maps of a Rat Brain Section}
\label{sec:ParameterMaps}

\begin{figure}[htbp]
\centering
\includegraphics[width=0.9 \textwidth]{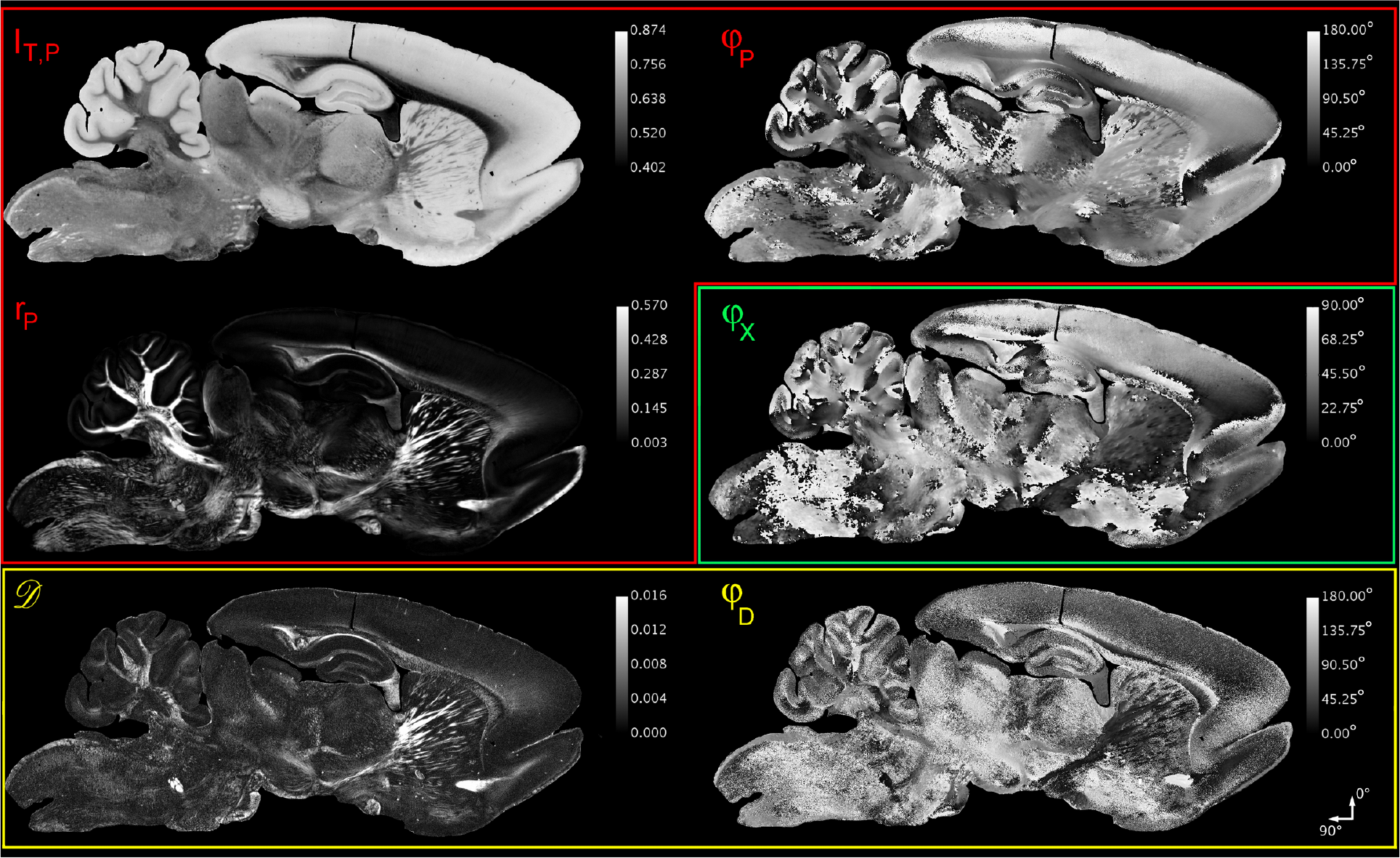}
\caption{Parameter maps of one rat brain section (s0175) obtained from the 3D-PLI (red), XP (green), and DI (yellow) measurements}
\label{fig:parametermaps}
\end{figure}

Figure \ref{fig:parametermaps} shows the parameter maps computed from the 3D-PLI measurement (transmittance $I_{T,P}$, direction angle $\varphi_P$, retardation $r_P$), the XP measurement (direction angle $\varphi_X$), and the DI measurement (diattenuation $\mathscr{D}$, direction angle $\varphi_D$) for one rat brain section (s0175).
 
\section{Uncertainties of the DI and 3D-PLI Direction Angles as a Function of the Retardation}
\label{sec:DeltaDIR_vs_RET}

\begin{figure}[htbp]
\centering $
	\begin{array}{ll}
	\textbf{(a)} & \textbf{(b)} \\
	\includegraphics[height=0.4\textwidth, trim={0 0 1.85cm 0}, clip]{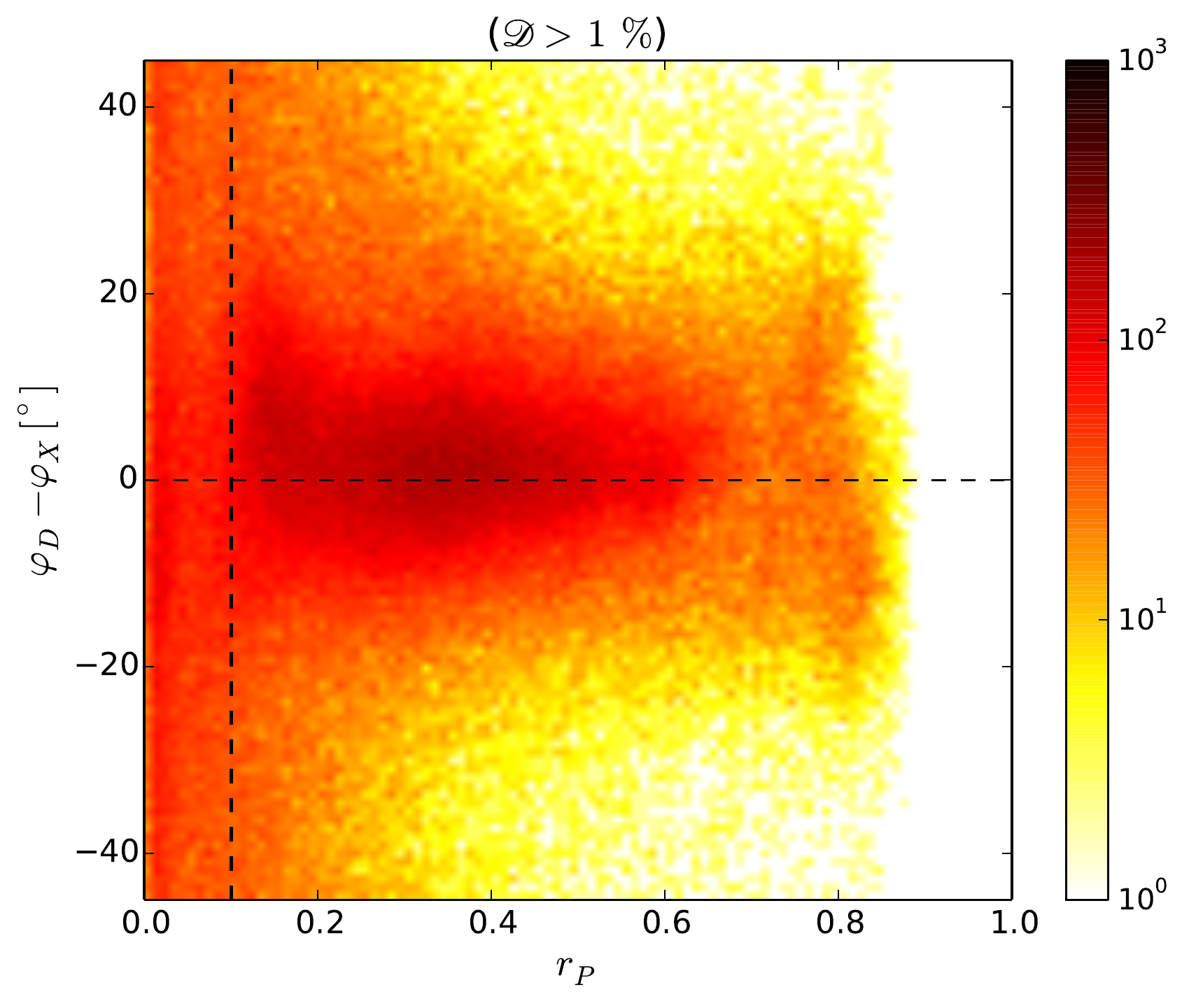} &
	\includegraphics[height=0.4\textwidth, trim={0 0 0 0}, clip]{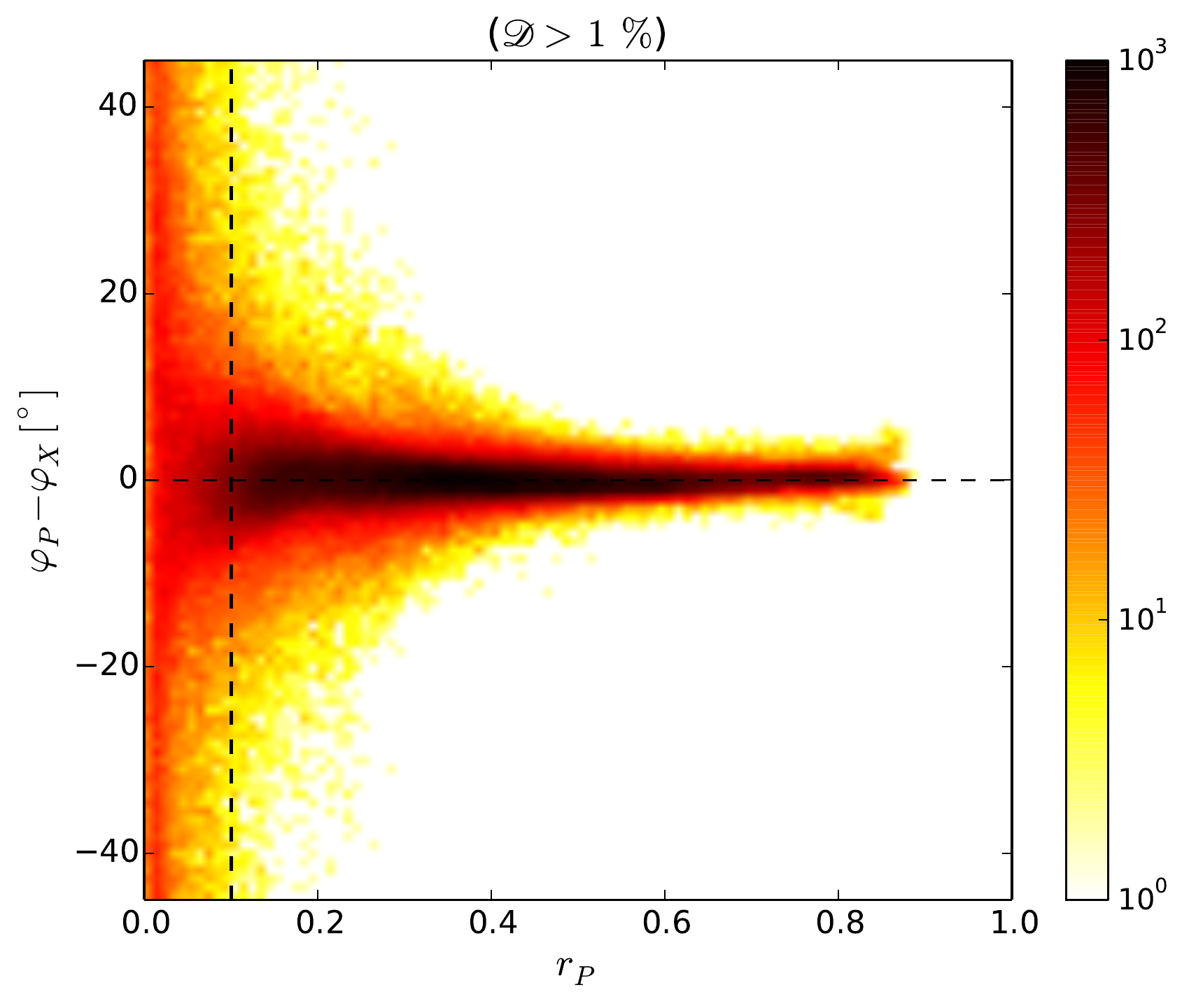} 
	\end{array}$
\caption{\textbf{(a)} 2D histogram showing the difference between the direction angle $\varphi_D$ derived from the DI measurement and the direction angle $\varphi_X$ derived from the XP measurement plotted against the measured retardation $r_P$. \textbf{(b)} 2D histogram showing the difference between the direction angle $\varphi_P$ derived from the 3D-PLI measurement and $\varphi_X$ plotted against $r_P$. The vertical dashed line marks the region ($r_P < 0.1$) for which the signal-to-noise ratio of $\varphi_X$ is low. The number of bins in the 2D histograms is 100 for both axes, respectively.}
\label{fig:DeltaDIR-XPvsRET}
\end{figure}

Figure \ref{fig:DeltaDIR-XPvsRET} shows the uncertainties of the direction angles $\varphi_D$ and $\varphi_P$ plotted against the measured retardation $r_P$ for all investigated brain sections and regions with $\mathscr{D}>1\,\%$.

The distributions of $(\varphi_D-\varphi_X)$ and $(\varphi_P-\varphi_X)$ are almost uniform for $r_P < 0.1$ (marked by the vertical dashed lines). This behavior can be explained by the low signal-to-noise ratio of $\varphi_X$ for small retardation values. Due to the $90^{\circ}$-orientation of the polarizers in the XP measurement, the transmitted light intensity approaches zero for small retardations (cf.\ \cref{fig:Setups}f). 
For larger retardation values, the signal-to-noise ratio of $\varphi_X$ increases and the distribution of $\varphi_D$ appears to be mostly independent of the retardation (it depends more on the diattenuation, see \cref{fig:DeltaDIR-XPvsDIA}a). This also agrees with the observation that retardation and diattenuation are not correlated for $\mathscr{D}>1\,\%$ (see \cref{fig:DIAvsRET}c). The distribution of $\varphi_P$ is much narrower than for $\varphi_D$ (in regions with $r_P > 0.1$) and also mostly independent of the retardation.

Based on these observations, the direction angles $\varphi_D$ and $\varphi_P$ were only evaluated in regions with retardation values $r_P > 0.1$ to avoid misinterpretation.


\section{List of Symbols and Abbreviations}
\label{sec:Symbols}

\setlength{\columnsep}{40pt}
\setlength\columnseprule{0.4pt}
\begin{multicols}{2}

\begin{tabbing}
XXXX, \= XXXXXXXXXXX \kill

$a_k$		\> Fourier coefficients associated with \\ \> sines\\[0.2cm]
ac			\> anterior commissure\\[0.2cm]
aci			\> anterior commissure intrabulbar part\\[0.2cm]
$\alpha$	\> out-of-plane inclination angle of the \\ \> fibers \\[0.2cm]
$\tilde{\alpha}$	\> modified inclination angle corrected \\ \> by the maximum measurable \\ \> retardation \\[0.2cm]
$\alpha_P$	\> inclination angle obtained from \\ \> 3D-PLI\\[0.2cm]
$b_k$		\> Fourier coefficients associated with \\ \> cosines\\[0.2cm]
C			\> index denoting the camera\\[0.2cm]
Cb			\> cerebellum\\[0.2cm]
cc			\> corpus callosum\\[0.2cm]
CCD			\> charge-coupled device\\[0.2cm]
cg			\> cingulum\\[0.2cm]
CPu			\> caudate putamen\\[0.2cm]
cu			\> cuneate fasciculus\\[0.2cm]
$\mathcal{C}$ \> image contrast\\[0.2cm]
$d$			\> sample/section thickness\\[0.2cm]
D			\> index denoting the DI meas.\\[0.2cm]
DI			\> Diattenuation Imaging\\[0.2cm]
$D$			\> diattenuation (of brain tissue)\\[0.2cm]
$D_D$		\> diattenuation obtained from DI\\[0.2cm]
$D_x$		\> diattenuation of the polarizer\\[0.2cm]
$D_y$		\> diattenuation of the analyzer\\[0.2cm]
$D^+$		\> diattenuation for which the axis of \\ \> maximum transmittance is parallel \\ \> to the fibers\\[0.2cm]
$D^-$		\> diattenuation for which the axis of \\ \> maximum transmittance is \\ \>  perpendicular to the fibers\\[0.2cm]
$\mathscr{D}$ \> measured diattenuation ($= D_D\, D_x$)\\[0.2cm]
$\delta$	\> phase retardation ($\approx \frac{2\pi}{\lambda} \, d \,\Delta n \,\cos^2\alpha$) \\[0.2cm]	
$\Delta$	\> sum of squared differences\\[0.2cm]
$\varphi$ 	\> phase; in-plane direction angle of the \\ \> fibers \\[0.2cm]
$\varphi_D$	\> direction angle from DI meas.\\[0.2cm]
$\varphi_P$	\> direction angle from 3D-PLI meas.\\[0.2cm]
$\varphi_X$	\> direction angle from XP meas.\\[0.2cm]
$\gamma$	\> retardance of the (quarter-)wave \\ \> retarder\\[0.2cm]
$I$			\> (total) intensity of light\\[0.2cm]
$I_{T,P}$	\> transmittance ($=$ average transmitted \\ \> intensity) \\[0.2cm]
$I_0$		\> intensity of the incident light \\[0.2cm]
$I_{\text{max}}$ \> maximum transmitted light intensity; \\
			\> maximum image intensity \\[0.2cm]
$I_{\text{min}}$ \> minimum transmitted light intensity; \\
			\> minimum image intensity \\[0.2cm]
$I_D(\rho)$	\> transmitted intensity in DI meas.\\[0.2cm]
$I_P(\rho)$	\> transmitted intensity in 3D-PLI meas.\\[0.2cm]
$I_X(\rho)$	\> transmitted intensity in XP meas.\\[0.2cm]
L			\> index denoting the light source\\[0.2cm]
LAP			\> Large-Area Polarimeter\\[0.2cm]
LED			\> light emitting diode\\[0.2cm]
$\varLambda$	\> Müller matrix of the (quarter-)wave \\ \> retarder\\[0.2cm]
$\lambda$ 	\> wave length\\[0.2cm]
$M$			\> Müller matrix of the brain tissue\\[0.2cm]	
$\mathcal{M}$	\> general Müller matrix\\[0.2cm]	
$\mu$		\> mean of a Gaussian distribution\\[0.2cm]
OCT			\> optical coherence tomography\\[0.2cm]
opt			\> optic tract\\[0.2cm]
P			\> index denoting the 3D-PLI meas.\\[0.2cm]
PLI			\> Polarized Light Imaging\\[0.2cm]
$P_x$		\> Müller matrix of the polarizer\\[0.2cm]
$P_y$		\> Müller matrix of the analyzer\\[0.2cm]
$p$			\> degree of polarization\\[0.2cm]
px			\> pixel\\[0.2cm]
$\psi$		\> spherical angle of Stokes vector \\ \> $\in [0,\pi]$ \\[0.2cm]
$\chi$		\> spherical angle of Stokes vector \\ \> $\in [-\pi/4,\pi/4]$ \\[0.2cm]
$n_e$		\> extraordinary refractive index\\[0.2cm]
$n_o$		\> ordinary refractive index\\[0.2cm]
$\Delta n$	\> birefringence ($n_E - n_o$)\\[0.2cm]
RNFL		\> retinal nerve fiber layer\\[0.2cm]
$R$			\> Müller matrix describing a rotation\\[0.2cm]
$r_P$  		\> retardation ($\vert\sin\delta_P\vert$) obtained \\ \> from 3D-PLI\\[0.2cm]
$r_{\text{max}}$  		\> maximum measurable retardation\\[0.2cm]
$\rho$ 		\> rotation angle of the polarizing filers \\[0.2cm]
$\vec{S}$	\> Stokes vector\\[0.2cm]
$\vec{S}_{\text{unpol}}$	\> Stokes vector for unpolarized light\\[0.2cm]
$\sigma$	\> standard deviation of a Gaussian \\ \> distribution\\[0.2cm]
$\tau$		\> average transmittance (of the brain \\ \> tissue) \\[0.2cm]
$\tau_x$	\> average transmittance of the polarizer \\[0.2cm]
$\tau_y$	\> average transmittance of the analyzer \\[0.2cm]
$\tau_{\Lambda}$	\> average transmittance of the retarder \\[0.2cm]
vhc			\> ventral hippocampal commissure\\[0.2cm]
WM			\> selected white matter regions\\[0.2cm]
X			\> index denoting the XP meas.\\[0.2cm]
XP			\> crossed polars\\[0.2cm]

\end{tabbing}
\end{multicols}


\end{document}